\documentclass{aa}  
\usepackage[utf8]{inputenc}
\usepackage{booktabs}
\usepackage{multirow}
\usepackage{siunitx}

\usepackage[T1]{fontenc}
\usepackage{lmodern}

\newcommand{\kms}{\,${\rm kms^{-1}}$ }
\usepackage{graphicx}
\usepackage{txfonts}
\usepackage[colorlinks,linkcolor={blue},citecolor={blue},urlcolor={blue}]{hyperref} 
\usepackage{comment}
\usepackage{color}
\usepackage{xcolor}
\definecolor{mhi}{rgb}{0.6,0.0,0.6}

\colorlet{avi}{green!0!orange!100!}

\begin{document} 

\titlerunning{FVSS--V. Mass modelling of the BCG NGC~1399 out to 150 kpc}
\authorrunning{A. Chaturvedi et al.}
\title{The Fornax Cluster VLT Spectroscopic Survey 
-- V. Mass modelling of the BCG NGC~1399 out to 150 kpc}

\author{Avinash Chaturvedi \inst{\ref{finca}, \ref{aip}, \ref{eso}}
\and Nicola R. Napolitano \inst{\ref{niunina}, \ref{inafnapoli}}
\and Michael Hilker       \inst{\ref{eso}}
\and Katja Fahrion        \inst{\ref{vienna}}
\and Sabine Thater        \inst{\ref{vienna}}
\and Glenn van de Ven     \inst{\ref{vienna}}
\and Michele Cantiello    \inst{\ref{inafteramo}}
\and Chiara Spiniello     \inst {\ref{eso}, \ref{inafnapoli}, \ref{oxf}}
\and Maurizio Paolillo    \inst{\ref{inafnapoli}, \ref{uninap}}
\and Tadeja Ver\v{s}i\v{c}        \inst{\ref{eso}, \ref{vienna}}}

\institute{
Finnish Centre for Astronomy with ESO (FINCA),
University of Turku, 20014 Turku, Finland
\label{finca}\\
\email{avinash.chaturvedi@utu.fi}
\and
Leibniz Institute for Astrophysics Potsdam (AIP),
An der Sternwarte 16, D-14482 Potsdam, Germany
\label{aip}
\and
European Southern Observatory,
Karl-Schwarzschild-Stra\ss{}e 2, 85748 Garching, Germany
\label{eso}
\and
Department of Physics ``E. Pancini'',
University of Naples Federico II,
Via Cintia 21, 80126 Naples, Italy
\label{niunina}
\and
INAF--Osservatorio Astronomico di Capodimonte,
Salita Moiariello 16, 80131 Naples, Italy
\label{inafnapoli}
\and
Department of Astrophysics, University of Vienna,
T\"urkenschanzstra\ss{}e 17, 1180 Vienna, Austria
\label{vienna}
\and
INAF--Astronomical Observatory of Abruzzo,
Via Maggini, 64100 Teramo, Italy
\label{inafteramo}
\and
Sub-Department of Astrophysics, Department of Physics,
University of Oxford, Denys Wilkinson Building,
Keble Road, Oxford OX1 3RH, UK
\label{oxf}
\and
University of Naples Federico II,
C.U. Monte Sant'Angelo, Via Cinthia,
80126 Naples, Italy
\label{uninap}
}

\date{Received: 15 May 2025/ Accepted: 31 July 2026}

% \abstract{}{}{}{}{} 
% 5 {} token are mandatory
 
\abstract 
{NGC~1399, the bright central galaxy of the Fornax cluster, contains an extensive population of globular clusters (GCs) extending beyond 200 kpc into its outer halo, approximately 6.5 effective radii ($\rm R_{e}$). In this paper, we conducted dynamical mass modelling of NGC~1399 out to 5 $\rm R_{e}$ ($\sim$150 kpc), using the most comprehensive radial velocity catalogue of GCs, which includes data from the Fornax cluster VLT spectroscopic survey (FVSS) and previous velocity measurements from the literature. Applying spherical Jeans equations, we performed dispersion-kurtosis modelling of NGC~1399's GC kinematics to derive its mass profile and the orbital anisotropy of its GCs. We also investigated the impact of including intra-cluster GCs in the mass modelling, which were selected based on spatial segregation and a velocity $\sigma$-clipping method. Our findings indicate that both cusp-like (NFW) and core-like (Burkert) halos can reproduce the observed kinematics of GCs. Including the intra-cluster GCs in the mass modelling yields broadly consistent mass profiles for both halo types, with the full GC sample giving a virial mass of $\log M_{\rm vir} = 13.81 \pm 0.09\,M_{\odot}$ for the NFW halo, slightly higher than the value inferred for the Burkert halo. Regardless of the dark matter halo profile adopted, we observe that the intra-cluster GCs exhibit radial anisotropy in the outskirts, especially among the blue, metal-poor GCs. These results suggest that GCs in NGC~1399's outer halo are influenced by an additional halo component associated with the galaxy cluster potential. The radial anisotropy observed in these outer-halo GCs provides strong dynamical evidence supporting their accretion from external sources, consistent with prior photometric and spectroscopic studies of GCs. This study highlights the importance of including intra-cluster GCs in dynamical mass modelling to fully account for the contribution of dark matter to the cluster gravitational potential.}


\maketitle

% For Introduction check introduction.tex
\section{Introduction}\label{intro}
Massive early-type galaxies (ETGs) are among the most luminous and massive systems in the Universe. They usually reside in the centres of galaxy clusters and represent the final stage of galaxy evolution. A prominent subset of these are the bright central galaxies (BCG), which sit near the bottom of the cluster potential well and often exhibit extended stellar halos and signatures of complex assembly histories. According to the $\Lambda$CDM cosmological structure formation scenario, ETGs are embedded in large dark matter (DM) halos and their growth is proposed to occur in a two-phase scenario. First, an intense burst of star formation forms a central spheroidal component in-situ, and in the second phase the outer parts are assembled ex-situ with the accretion or merger of lower mass galaxies \citep{Oser2010, Dokkum2010}. 

The entire assembly of the massive DM halo occurs in a hierarchical way through the merging of smaller DM halos \citep{Cooper2013, Pillepich2015}. The hierarchical formation of ETGs leaves an imprint on their structures and physical properties, leading to the build-up of diffuse stellar light at large scales \citep{Gregg1998, Napolitano2003,Montes2014, Arnaboldi2012, Iodice2019} and intra-cluster stellar populations such as accreted globular clusters (GCs) in the outskirts of galaxy halos \citep{Harris2020, Longobardi2018, Cantiello2020, Chaturvedi2022, Arnaboldi2022}. This kind of evolution of ETGs makes them important systems for understanding galaxy evolution and formation at the largest and most massive scales. However, measuring the total mass profile on a large physical scale, particularly of the extended DM halo of ETGs, is pretty challenging. Unlike late-type galaxies, like spirals and dwarf irregulars, ETGs are dominated by old stellar populations. Moreover, the stellar light in the outskirts of the ETGs is very faint, making it difficult to perform a detailed study. 

In recent years, with technological advancement, there has been a great interest in comprehending the nature and origin of massive galaxies using the diffuse intra-cluster light (ICL) in groups and clusters of galaxies through observations \citep{Iodice2016, Iodice2019, Kluge2020} and cosmological simulations \citep{Napolitano2003,Cooper2015, Pillepich2015, Marini2022}. The ICL is considered an important tracer for measuring the gravitational potential of the system, as it extends out to several 100 kpc, which is similar to the size of DM halos, and thus provides a direct opportunity to understand the mass assembly and DM content of ETGs \citep{Montes2018, Montes2019, Montes2022}. However, due to the low surface brightness nature of the ICL (mostly $\mu_g > 29$ mag arcsec$^{2}$), it is very challenging to obtain spectroscopy and any kinematical information directly. An alternative is to use discrete kinematical tracers such as GCs and planetary nebulae, which are observable out to tens of effective radii ($\rm R_{\rm e}$) \citep[see][for a review]{Arnaboldi2022}.

GCs are bright, compact sources and are excellent kinematic tracers of the galaxy potential. In broad terms, the GCs population, especially in the bright massive galaxies, can be split into two classes, metal-rich (red) GCs and metal-poor (blue) GCs \citep{Brodie2006}. These two classes show different kinematical properties and spatial distributions. The red GCs follow the stellar light distribution of their host galaxy and have kinematics similar to that of the galaxy's inner spheroid (see e.g. \citealt{napolitano2014}). In contrast, the blue GCs dominate in the outskirts of galaxy halos and have complex kinematics \citep{Schuberth2008, Coccato2013, Pota2018, Chaturvedi2022}.The observed velocity dispersion profiles of red and blue GCs also differ, with blue GCs showing slightly higher values than their red counterparts \citep{Chaturvedi2022}. The different properties of GCs suggest different origins of the two classes, with the red GCs born in-situ and the blue GCs being accreted from infalling low-mass cluster galaxies \citep{Ashman1992, Kundu2001, Peng2006}. There is evidence that these properties are related to the two-phase hierarchical formation scenario of galaxy formation \citep{Cooper2013, Pillepich2015}. 

A key challenge in dynamical mass modelling is the mass–anisotropy degeneracy, which arises from the difficulty in determining whether the observed variation in velocity dispersion of tracers is due to orbital anisotropy or the underlying mass distribution. The distinct kinematic behaviour and spatial distribution of the red and blue GCs allows to break the mass-anisotropy degeneracy. In addition, the availability of kinematic tracers out to large physical scales alleviates the degeneracy between dynamical stellar mass-to-light ratio (M/L) and DM mass \citep{napolitano2009, napolitano2014, Zhu2016, Chao2020, Tadeja2024jan}. Furthermore, studies of extended planetary nebula kinematics around NGC 1399 (\citealt{Napolitano2002}) and of the extended stellar light in the Hydra\,I cluster core (\citealt{Hilker2018}) revealed that kinematic substructures can boost the line-of-sight velocity dispersion, and influence the mass-modelling results. Other similar works point toward the same picture that ignoring the presence of the substructures in the mass modelling affects the dynamical mass estimates \citep{Old2017, Tucker2020}.

After the Virgo cluster, the Fornax galaxy cluster, at $\sim$ 20 Mpc \citep{Blakeslee2009} is the nearest galaxy cluster and its central galaxy, NGC\,1399, is a well studied system that hosts a large number of GCs. It provides an excellent opportunity to investigate the formation and evolution of an ETG in a dense environment. Previous photometric studies of the Fornax cluster have detected the presence of GC substructures between NGC\,1399 and its neighbouring galaxies \citep{Abrusco2016, Iodice2019, Cantiello2020}, and therefore the Fornax cluster offers the chance to investigate the role of subtructures on the mass-modelling.

To understand the mass-assembly of the Fornax cluster, we are conducting the Fornax cluster VLT spectroscopic survey \citep[hereafter FVSS, for details, see][]{Pota2018, Spiniello2018, Chaturvedi2022, Napolitano2022} targeting GCs and planetary nebulae in the central 1.5 square degrees. FVSS is a multi-instrument observing campaign using FORS2, VIMOS, and FLAMES at the ESO VLT (see \citealt{Pota2018}). \citet[][hereafter FVSS-III]{Chaturvedi2022}, combined the FVSS VLT/VIMOS results with previous GCs velocity measurements mainly from the Fornax 3D survey \citep{Katja2020} and \cite{Schuberth2008} and produced the largest GCs radial velocity catalogue of more than 2300 GCs in the Fornax cluster. In FVSS-III, we provided the spectroscopic confirmation and kinematical characterization of the GC substructures between the Fornax cluster member galaxies. We also showed that the outer halo of NGC\,1399 and intra-clusters substructures are dominated mainly by blue GCs. The photometrically detected GC streams in Fornax were also confirmed statistically in phase space using the COSTA algorithm \citep{Gatto2020} in  \cite{Napolitano2022}.

In the present work, we use the kinematics of the GC systems around NGC\,1399, the BCG of the Fornax cluster, to obtain its mass profile out to 5 $\rm R_{e}$ ($\sim$ 150 kpc). We perform a dispersion-kurtosis Jeans modelling, adopted from \citet{napolitano2014}, to reduce the mass-anisotropy degeneracy and obtain constraints on the orbital distribution of the red and blue GC populations. The key goals of this work are a) to obtain a total mass estimate of NGC\,1399, b) to investigate the effect of the substructures and intra-cluster GCs on the mass modelling, and c) to gain insight into the orbital distribution of the GCs. 

This paper is organized as follows: In Section \ref{sec2} we present the spatial density and kinematics of the GCs. Section \ref{sec3} briefly introduces the Jeans dynamical modelling and fitting method, and in Section \ref{sec4} we present the results obtained from the modelling work. Finally, in Section \ref{sec5}, we present the summary and conclusion of our work. 

%################ SECTION 2 ########### 
\section{Data}\label{sec2}

This section presents the observables required for the Jeans modelling work of Section \ref{sec3}, including the GCs surface density profiles and kinematics details of the GCs sample. As mentioned in the introduction, we use the GCs radial velocity catalogue presented in FVSS-III and additional literature compilation for the mass-modelling work of NGC~1399. For a detailed description of the data reduction and radial velocity measurements, we refer to paper FVSS-III.

\subsection{GCs sample for dynamical modelling} \label{sec2:sample}

% FIGURE 1
\begin{figure*}
\centering
\sidecaption
\includegraphics[width=18cm]{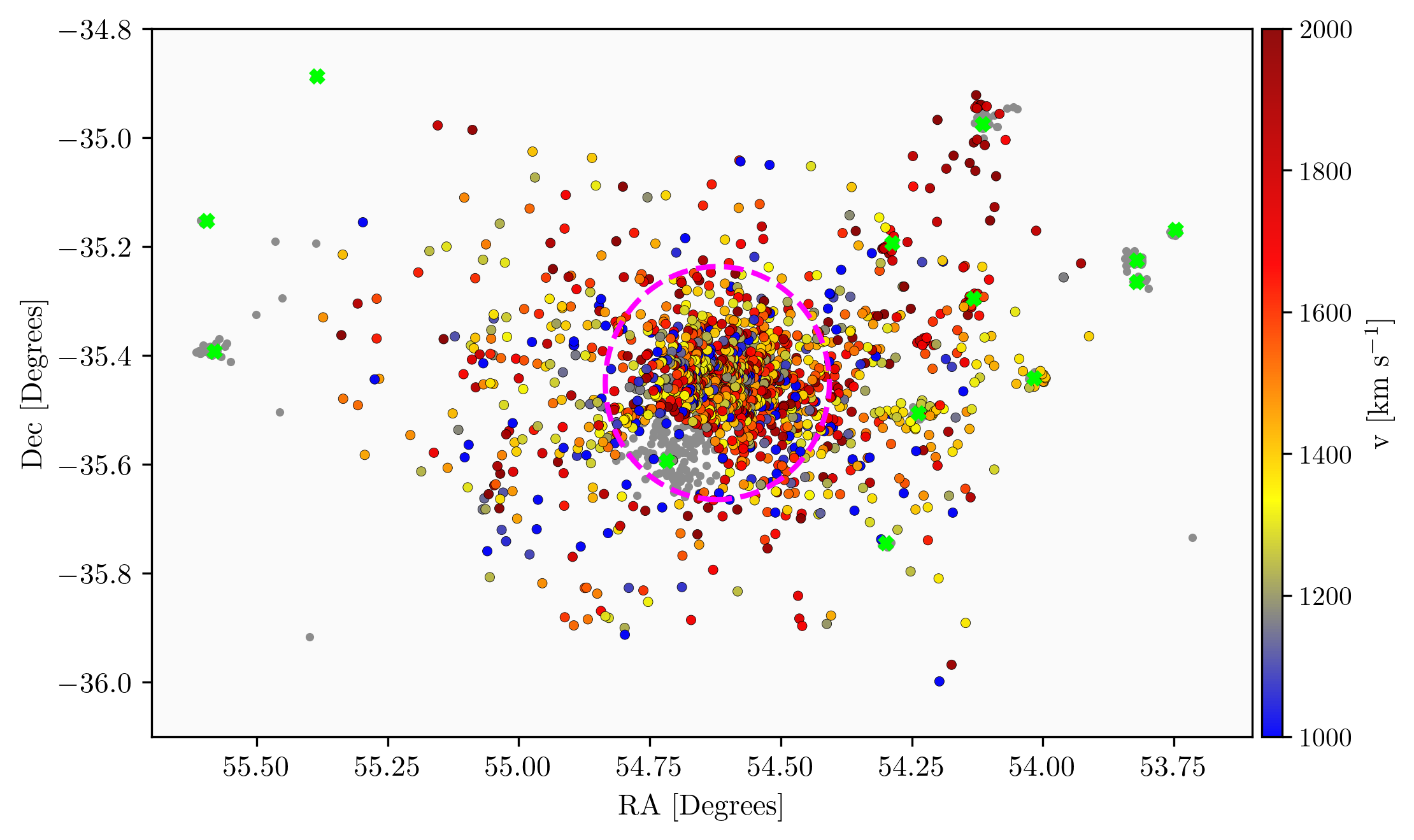}
\caption{Radial velocity map of the \textit{inner} and \textit{full} GC samples of NGC~1399, including the IC component, within 1.5 square degrees of the Fornax cluster. Grey points show the full FVSS-III catalogue (2341 GCs). Coloured points represent the \textit{inner} and \textit{full} samples, colour-coded by their line-of-sight velocities. The dashed magenta circle marks the radial extent of the \textit{inner} sample ($\rm R \leq 2.5\,R_e$). Lime crosses indicate the positions of major Fornax cluster galaxies. }
\label{fig:rad_map}
\end{figure*}

% FIGURE 2
\begin{figure}
\centering
\includegraphics[width=1.05\linewidth]{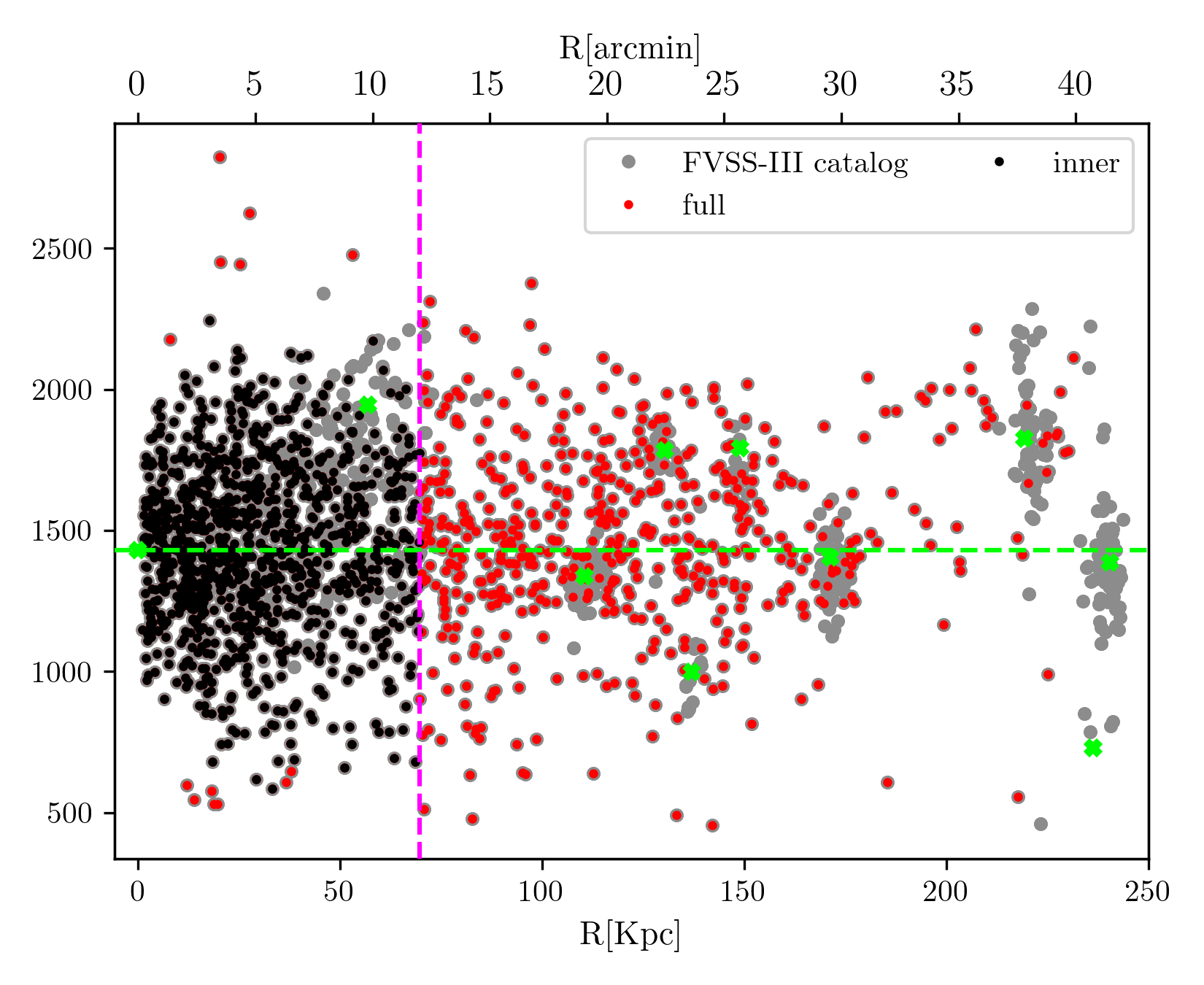}
\caption{Line-of-sight velocities of the GC sample (see Section \ref{sec2}) plotted as a function of cluster-centric distance. Grey points represent the full GC catalogue from the FVSS-III study. Black points show the \textit{inner} GC sample and together with the red points they from the \textit{full} GC samples, respectively. Lime crosses indicate the positions of major galaxies, and the horizontal dashed line marks the systemic velocity of NGC~1399. Grey points visible around the lime crosses correspond to GCs that are excluded following the radial sigma-clipping procedure described in Section \ref{sec2:sample}. The vertical dashed magenta line indicates $\rm 2.5\,R_e$ of NGC~1399, corresponding to the radial cut used to define the \textit{inner} sample besides the 2.5$\sigma$ GCs LOS scatter around NGC 1399.}
\label{fig:proj_los}
\end{figure}

\begin{figure*}
    \centering
    \includegraphics[width=0.85\linewidth]{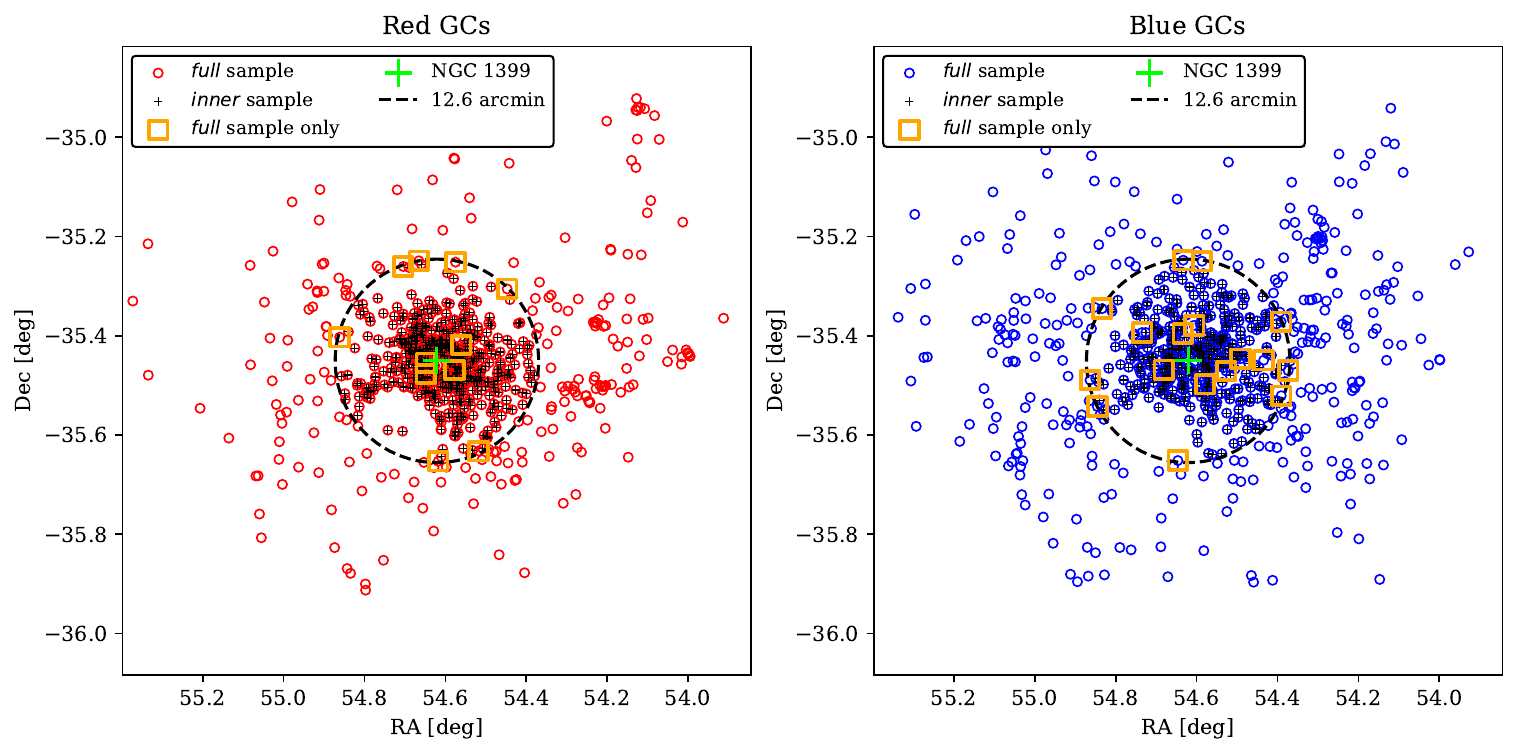}
    \caption{Spatial distributions of the \textit{inner} and \textit{full} samples, separated into the red (left panel) and blue (right panel) GC subpopulations. In both panels, the black dashed circle marks the radial boundary of the \textit{inner} sample, while the orange squares indicate GCs that are present in the \textit{full} sample but not in the \textit{inner} sample.}
    \label{fig:spatial_gc_samples}
\end{figure*}

One of the primary goals of this paper is to assess the influence of intra-cluster GCs (hereafter ICGCs) on the mass-modelling of NGC~1399. An accurate definition of the intra-cluster population should involve dynamical arguments about how bound a population of test-particles is to the galaxy potential \citep{Dolag2010}. This would imply robustly separating the central galaxy potential from the cluster potential and selecting as bound members those that are solidly inside the escape velocity of the central galaxy potential with respect to the cluster potential. 

In this work, we take a first step toward a more compelling separation by evaluating the impact of a non-rigorous but conservative ICGCs population, defined based on a more heuristic approach. For this purpose, we define two samples of GCs for our modelling work. The first sample consists of GCs within 2.5$\rm R_{e}$ of NGC~1399, where we exclude extreme radial velocity GCs (hereafter referred to as \textit{inner} sample) that are likely unbound to the galaxy and are just projected onto its central region by chance. The second sample consists of the entire sample of GCs, including the ones in the outer halo of NGC~1399 extending out to 5$\rm R_{e}$ (hereafter referred to as \textit{full} sample). This sample includes, besides the \textit{inner} sample, also: 1) a dominant component of GCs belonging to the intra-cluster component, and 2) the GCs bound to to all other galaxy cluster members in the area covered by the FVSS sample.

For the \textit{inner} sample, we selected the GCs by performing cuts in a phase-space distribution, similar to paper FVSS-III. There, we first calculated the scatter in the radial velocities of GCs within 2$\rm R_{e}$ of a galaxy and used $\pm$ 2 $\rm \sigma$ of this velocity scatter around the galaxy LOS as the lower and upper boundary to select the GCs belonging to each galaxy (for details, see section 5.2 of FVSS-III). In contrast to FVSS-III, instead of applying fixed cuts in velocity, we calculate the radially varying LOS velocity scatter of GCs within 2.5$\rm R_{e}$ of NGC~1399. We used a slightly extended radial cut compared to the previously used value of 2.0$\rm R_{e}$, as the latter was a strict cut that could potentially exclude the GCs belonging to outer regions of NGC~1399. We adopted the r band 1$\rm R_{e}$=5.13 arcminutes ($\sim$ 29.77 kpc) measured using the isophotal fitting from \cite{Iodice2019} \footnote{ Note that this effective radius has been obtained from the full photometry of NGC~1399, out to $\sim 190$ kpc, i.e. including the contribution of the intracluster diffuse light. As such, this has to be considered an upper limit of the effective radius of the ``bound'' stellar component, that we mean to isolate in the kinematical clipping.}. We used $\pm2.5\sigma$ velocity scatter around NGC~1399 LOS velocity as the lower and upper boundary to select the GCs belonging to the central galaxy.   

The giant elliptical NGC\,1404 lies at a projected distance of $2\rm R_{e}$ from NGC~1399, and its outer halo GCs overlap in phase space with those of NGC~1399. Therefore, we removed the GCs belonging to NGC\,1404 from the GC sample selected within 2$\rm R_{e}$. The GC selection around NGC\,1404 was carried out in a similar way as for the inner sample, but using the $\rm R_{e}$ of NGC\,1404 and a $\pm3\sigma$ velocity cut based on its own GCs. We adopted the $\pm3\sigma$ threshold in velocity space to ensure the effective removal of all GCs associated with NGC\,1404.

For the \textit{full} sample, we considered the entire FVSS-III catalogue of GCs but limiting it to a distance of 250 kpc due to the spatial incompleteness of the spectroscopic sample. Within this spatial range, the GC systems of six major galaxies—NGC\,1379, NGC\,1380, NGC\,1381, NGC\,1387, NGC\,1380B, and NGC\,1389—also contribute to the sample. To remove the GCs associated with these galaxies, we adopt an 2.0$\rm R_{e}$ cut in the phase-space and $\pm$3$\rm \sigma$ LOS velocity scatter of GCs for each galaxies in the velocity space. After removing the GCs from individual galaxies, this left a total of 1640 GCs for the \textit{full} sample. 

Figure \ref{fig:rad_map} presents the radial velocity map of the \textit{full} GC sample, while Figure \ref{fig:proj_los} displays the projected LOS velocity for the two different GC samples. Black and red dots represent the \textit{inner} and \textit{full} sample GCs, respectively. One can clearly notice the presence of GCs between the Fornax cluster galaxies. Despite the distinct separation of these two samples, we remark that the \textit{full} sample extends outside a radius of 10$'$, which has been found to mark a transition in the observed profiles of the light distribution \citep{Iodice2016}, X-ray distribution \citep{Paolillo2002}, mirrored by the dynamical properties \citep{Napolitano2002} and kinematics \citep{Spiniello2018} of the planetary nebulae  and GCs \citep{Pota2018}. All these evidences converge toward a scenario where the intra-cluster population is expected to dominate in the regions at $R>10'$ ($\sim$ 58 kpc), consistent with our selection. Figure~\ref{fig:spatial_gc_samples} shows the spatial distributions of the red and blue GCs in the \textit{inner} and \textit{full} samples. The photometric division used to separate the GCs into red and blue subpopulations is described in Section~\ref{sec2:tracerdensity}. The left and right panels show the distributions of the red and blue GCs, respectively. In both panels, the black dashed line marks the radial boundary of the \textit{inner} sample. The blue GCs appear to be more spatially extended and more uniformly distributed throughout the IC region. The orange squares in both panels indicate GCs that are present only in the \textit{full} sample but absent from the \textit{inner} sample. This difference arises because the sample selection is based on both projected phase space and LOS velocity. In addition to the phase-space cut, we applied a $\pm3\sigma$ LOS-velocity criterion to the GC system associated with each galaxy, except for NGC~1399, for which we adopted a $\pm2.5\sigma$ criterion. The GCs marked by the orange squares have large LOS velocity offsets relative to the systemic velocity of NGC~1399.

\begin{figure}[h]
    \centering
    \includegraphics[width=1.0\linewidth]{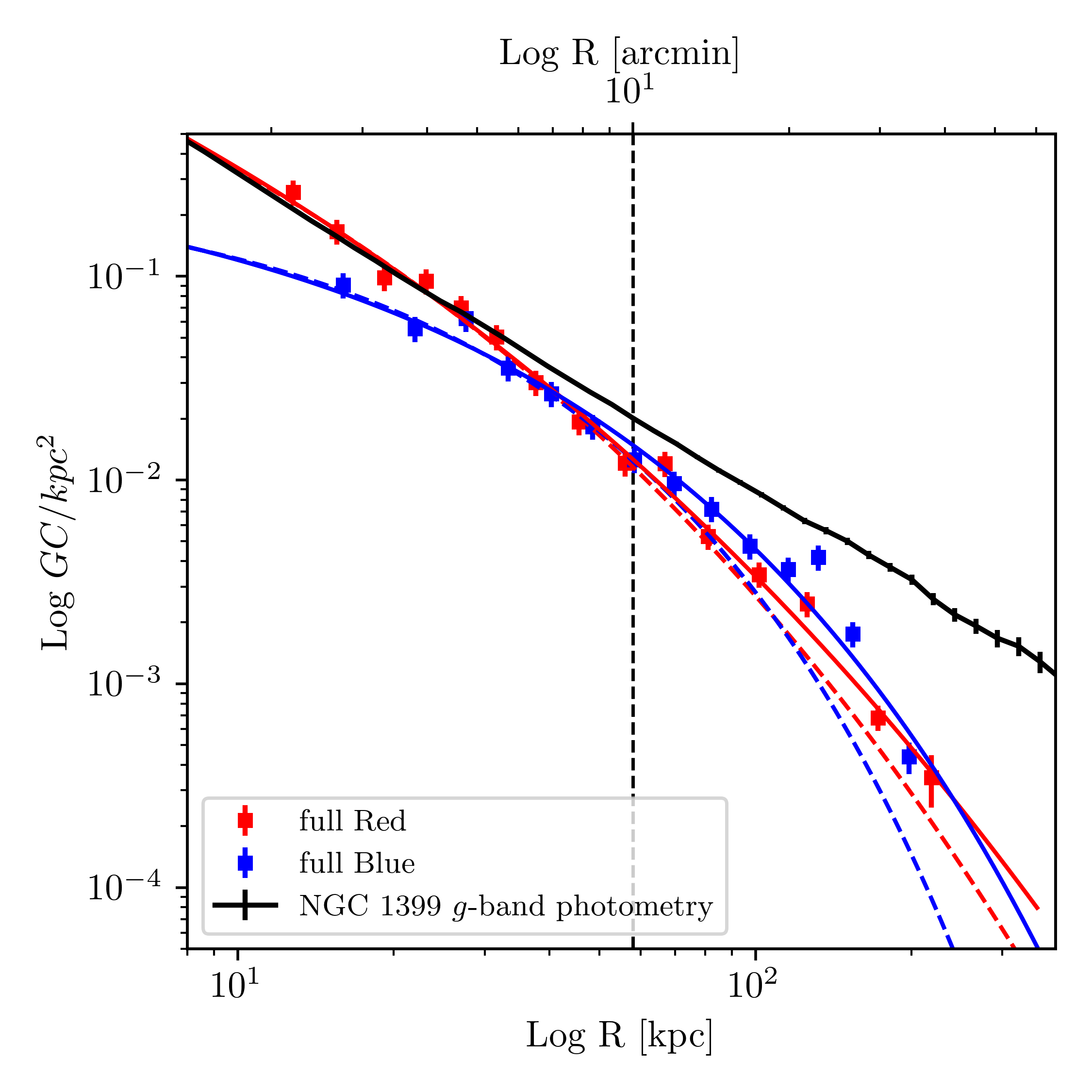}
    \caption{GCs radial surface density profiles around NGC~1399. The vertical dashed line indicates 1 $\rm R_{e}$ of NGC~1399. Blue and red squares indicates the observed radial surface density of blue and red GCs of the \textit{full} sample, respectively. The black curve shows the $g$ band scaled stellar surface brightness of NGC~1399 taken from \cite{Iodice2016}. Continuos red and blue curves represent the S\'{e}rsic fits to the red and blue GCs of the \textit{full} sample, respectively, while the dashed curve corresponds to the fit to the \textit{inner} sample.}
    \label{fig:gcs_nden}
\end{figure}

\subsection{GCs surface density} \label{sec2:tracerdensity}

As an input for the dynamical modelling (described in section \ref{sec3}), we need the GC tracer density. We selected GCs  brighter than i $<$ 23.2 mag to ensure a uniform sample of GCs magnitudes. To divide them into red and blue subpopulations, we adopted a $g-i$ colour value $\sim$ 0.97 as already done in FVSS-III. 

To measure the GC surface density, we divided both the \textit{inner} and \textit{full} spectroscopic GC samples into circular radial bins, each containing 50 GCs, and assumed Poissonian uncertainties. The assumption of circular symmetry is motivated by the classification of NGC~1399 as an E1 system \citep{Vaucouleurs1991}. However, more recent photometric studies have shown that NGC~1399 is not strictly spherical, with its outer regions exhibiting variations in ellipticity that may indicate an intrinsically triaxial structure \citep{Li2011, Pulsoni2018}. Mild flattening of the GC system has also been reported \citep{Cantiello2020}. To assess the impact of this on our results, we repeat the analysis and dynamical modelling in which all the GCs observables including surface density and kinematics were extracted using elliptical annuli that account for the observed flattening of the GC system. The results of this test are presented in Section~\ref{sec5.2}, where we show that the inferred mass and anisotropy profiles are not significantly affected by the assumed symmetry.

To account for the inner spatial incompleteness we fit the GCs surface density profile from 1 to 30 arcminutes and drop the first binned datapoint, which is below 1 arcmin. Similar to the work of \cite{Pota2015}, we characterised the GCs surface density profile, using the S\'{e}rsic function \citep{Sersic1963}, of the form:

\begin{equation} \label{equ1}
    I(r) = I_{o}\mathrm{exp}\Bigg\{-b\Bigg[\Bigg(\frac{R}{R_{e}}\Bigg)^{1/n_{e}}-1\Bigg]\Bigg\}, 
\end{equation} 
where $b = 1.9992 n - 0.3271$ and $I_{o}$ is the surface density at the characteristic radius $R_{e}$.

% ############# TABLE 1 #####################
\begin{table}[h]
\caption[]{Best fit parameters of the S\'{e}rsic profile for the GCs surface density profile.}
\renewcommand{\arraystretch}{1.15}
\label{Tab1:photo_param}
\centering
\begin{tabular}{ccccccc}
\hline

\textbf{Sample} & $I_{e}$ & $R_{e}$ & $n_{e}$  \\ 
\textit{inner}   & log [$GC/kpc^{2}$]  & [kpc]   \\ 
\hline
    % All   & 0.063 & 40.39 & 2.11  \\
    Red   & 0.075 & 24.51 & 3.33  \\
    Blue  & 0.025 & 41.17 & 1.41  \\

\hline
\textit{full}   &  &    \\ 
\hline
    % All   & 0.038 & 56.03 & 3.38  \\
    Red   & 0.058 & 27.83 & 4.35  \\
    Blue  & 0.015 & 56.09 & 1.83  \\
\hline
\end{tabular}
\end{table}

The best-fit values of the red and blue GCs populations of the \textit{inner} and \textit{full} samples are shown in Table \ref{Tab1:photo_param}. 

Figure \ref{fig:gcs_nden} shows the GCs number density profiles of the \textit{full} sample for red and blue GCs. The continuous lines indicate the S\'{e}rsic fits of the \textit{full} sample while the dashed curve indicates the fit for the \textit{inner} sample. For clarity, we display only the S\'{e}rsic fit and omit the data points of the \textit{inner} sample. Figure \ref{fig:gcs_nden} also shows the $g$-band stellar surface brightness of NGC~1399, adapted from the work of \cite{Iodice2016}. We find that the surface density profile of the red GCs closely follows the stellar surface brightness profile of the galaxy within the inner 3 kpc.

The projected number density profile can be used to derive the 3D density profile, needed for the dynamical modelling %\citep[described in Section \ref{sec3}, see also ][]{napolitano2014}.
described in Section \ref{sec3}.
The de-projection of the GC number density $j(r)$, is performed, assuming a spherical symmetry, via the Abel integration \citep{Binney1987}:

\begin{equation} \label{equ2}
j(r) = -\frac{1}{\pi}\int_{R}^{\infty}\frac{dI}{dR}\frac{dR}{\sqrt{R^{2}-r^{2}}} 
\end{equation}

where, $r$ and $R$ are the 3D and 2D radii respectively. $I(R)$ denotes the projected number density of the tracer, expressed in units of number/$\rm kpc^{2}$ for GCs. 

% FIGURE 4
\begin{figure*}[h]
    \centering
    \includegraphics[width=1.0\linewidth]{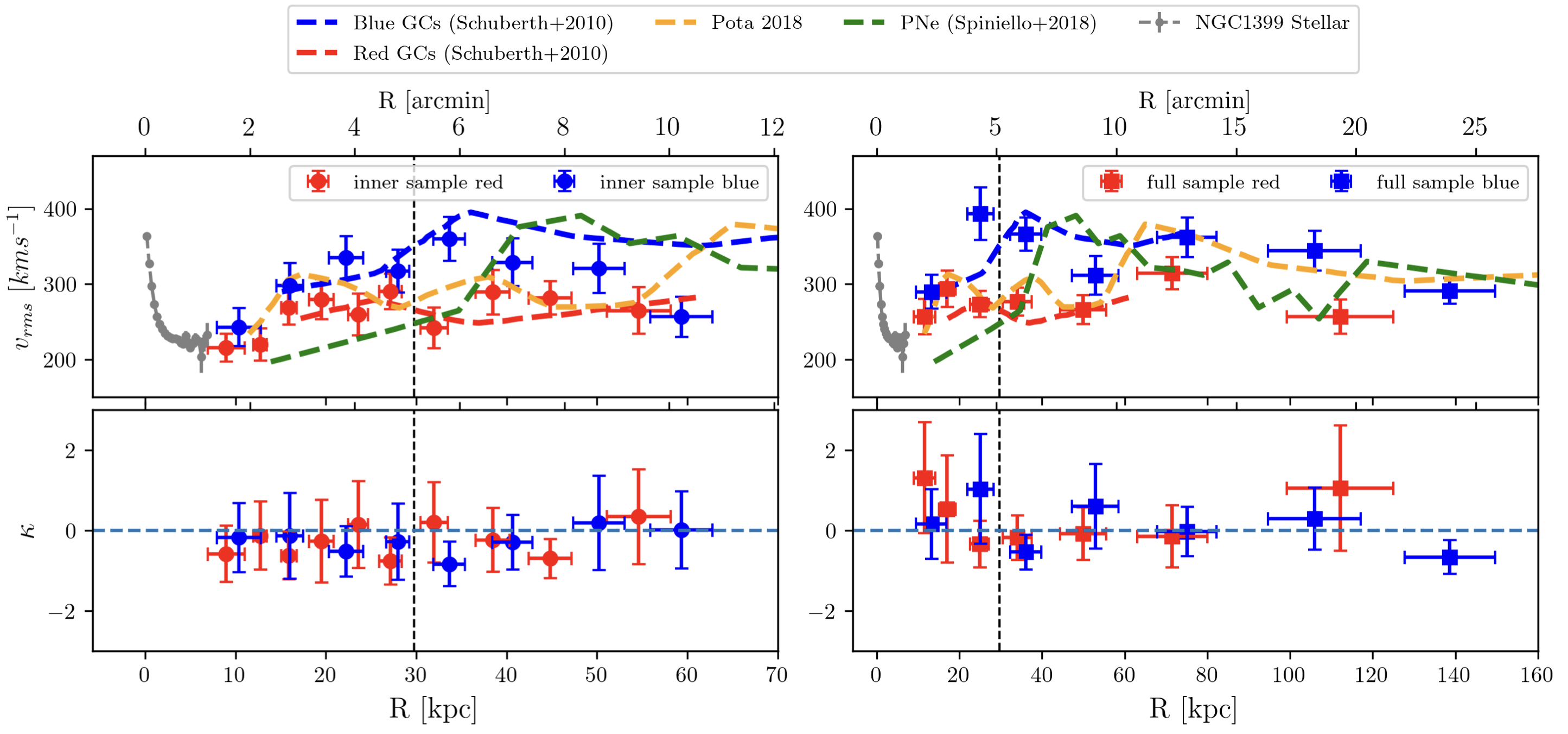}
    \caption{Velocity dispersion (top panels) and kurtosis (bottom panels) profiles of the two samples defined in Section \ref{sec2}. Left: the red and blue dots show the profiles for the red and blue GCs of the \textit{inner} sample. Right: same as the left panels, but for the \textit{full} sample. In the top panels, we also show velocity dispersion measurements from previous GC studies: \cite{Schuberth2010} (dashed blue and red lines for blue and red GCs), \cite{Pota2018} (dashed orange line), planetary nebulae from \cite{Spiniello2018} (dashed green line), and stellar kinematics of NGC~1399 from \cite{Katja2020} (grey dots).}
    \label{fig:velo_disp_kurt_pro}
\end{figure*}

\subsection{GCs kinematics}

The line of sight velocity distribution (LOSVD) of particles in a given potential is usually described, in the most general case, via Gauss Hermite polynomials \citep{Ortwin1993}. The first and second-order velocity moments quantify the Gaussian-like component of the LOSVD, and the higher-order moments measure the deviation of the LOSVD from a Gaussian distribution. Higher order moments such as kurtosis ($\kappa$) contain information about the orbital distribution of the GCs \citep{napolitano2014}. Following \cite{napolitano2014} we model the projected velocity dispersion and kurtosis profile using the high-order radial Jeans equations \citep[see also][]{Lokas2005}, as we detail in Section \ref{sec3}.

We calculated the GCs velocity dispersion in a similar manner as described in FVSS-III. We derived the velocity dispersion in each radial bin using the following expression:

\begin{equation} \label{equ3}
    v_{rms}^{2} = \frac{1}{N}\sum_{i=1}^{N} (v_{i} - v_{sys})^{2} - (\Delta v_{i})^{2}.
\end{equation}

Here, $v_{i}$ represents the radial velocity of the $i$th GCs, and $\Delta v_{i}$ denotes its velocity uncertainty. For the systemic velocity, we adopt a value $v_{sys}$ = 1430 \kms from \cite{Ferguson1989}. The uncertainty in the velocity dispersion is estimated using a bootstrapping technique. Specifically, for each radial bin, we compute the velocity dispersion 1000 times and take the 1$\rm \sigma$ scatter as the uncertainty. For the \textit{inner} sample, we used a bin size of 50 GCs for both red and blue subpopulations, whereas, for the \textit{full} sample, we used 100 GCs per bin. We verify that reducing the bin size does not affect the resulting dispersion profile for either sample. Although smaller bins introduce more noise into the profile, the overall trend remains unchanged.

In Figure \ref{fig:velo_disp_kurt_pro}, the top left panel shows the velocity dispersion profiles of the red and blue GCs of the \textit{inner} sample, and the right panel for the \textit{full} sample. The vertical dashed line indicates 1$\rm R_{e}$ of NGC~1399. We observe that, in both samples, the blue GCs have a higher velocity dispersion than the red GCs. For the \textit{inner} sample, the velocity dispersion profile appears smooth because the GCs were selected using a $\pm2.5\sigma$ LOS velocity cut around NGC~1399, which removes the extreme radial velocity GCs.  In contrast, for the \textit{full} sample, the velocity dispersion of blue GCs is, on average, approximately 100 \kms higher than that of the red GCs. 

In FVSS-III, we carried out a detailed comparison of the GC velocity dispersion with literature measurements from other kinematic tracers, such as planetary nebulae and previous GC studies; we refer the reader to that work for further details (Section 5.2). Here we briefly compare GCs  kinematics of our sample with the previous kinematics tracers in NGC~1399 and  the Fornax cluster. In Figure \ref{fig:velo_disp_kurt_pro}, we show velocity dispersion profiles from the literature, including GCs from \cite{Schuberth2010, Pota2018}, PNe from \cite{Spiniello2018}, and stellar kinematics from \cite{Katja2020}. We find that the \textit{inner} sample of red GCs is in good agreement with the stellar kinematics at around $\sim$8 kpc. At larger radii, the GC velocity dispersion follows the trends observed for PNe and for previously measured blue and red GC populations. We also note that the PNe velocity dispersion is in very good agreement with the \textit{inner} sample of red GCs. The velocity dispersion profiles of red and blue GCs from \cite{Schuberth2010} provide a good match to our sample. The \cite{Pota2018} sample (orange dashed line) follows the red GCs in the inner region, where red GCs dominate in number, while in the outer regions it agrees with our blue GC measurements for the \textit{full} sample. The PNe velocity dispersion (green dashed line) from \cite{Spiniello2018} varies between the red and blue GC values but remains consistent within the uncertainties.

Next, we measured the reduced kurtosis denoted as ($\kappa$) in the same circular radial bins as used for the velocity dispersion profiles with the following expression \citep{Joanes1998}:

\begin{equation} \label{equ4}
    \kappa = \frac{{\mu^{4}}}{\sigma^{4}} - 3
\end{equation}

Here $\mu^{4}$ and $\sigma$ denotes the fourth central moment and standard deviation, respectively. For a Gaussian LOSVD, $\kappa\sim0$, while a value of $\kappa>0$ suggests radial orbits, and $\kappa<0$ tangential orbits. The bottom panels of Figure \ref{fig:velo_disp_kurt_pro} show the GCs kurtosis profiles of both samples. For the \textit{inner} sample, the blue and red GCs have a relatively constant $\kappa$ close to $0$, possibly indicating isotropic or mildly tangential orbits.  In contrast, for the \textit{full} sample, the kurtosis profile shows significant noise within 1$\rm R_{e}$, with no clear distinction between the two populations. The difference between the kurtosis profiles of the \textit{inner} and \textit{full} samples over their overlapping radial range ($R \lesssim 60$ kpc) is mainly driven by a small number of GCs with extreme line-of-sight velocities that are included only in the \textit{full} sample, as indicated by the orange squares in Figure~\ref{fig:spatial_gc_samples}.

% FIGURE 5
\begin{figure}
\centering
\includegraphics[width=1.0\linewidth]{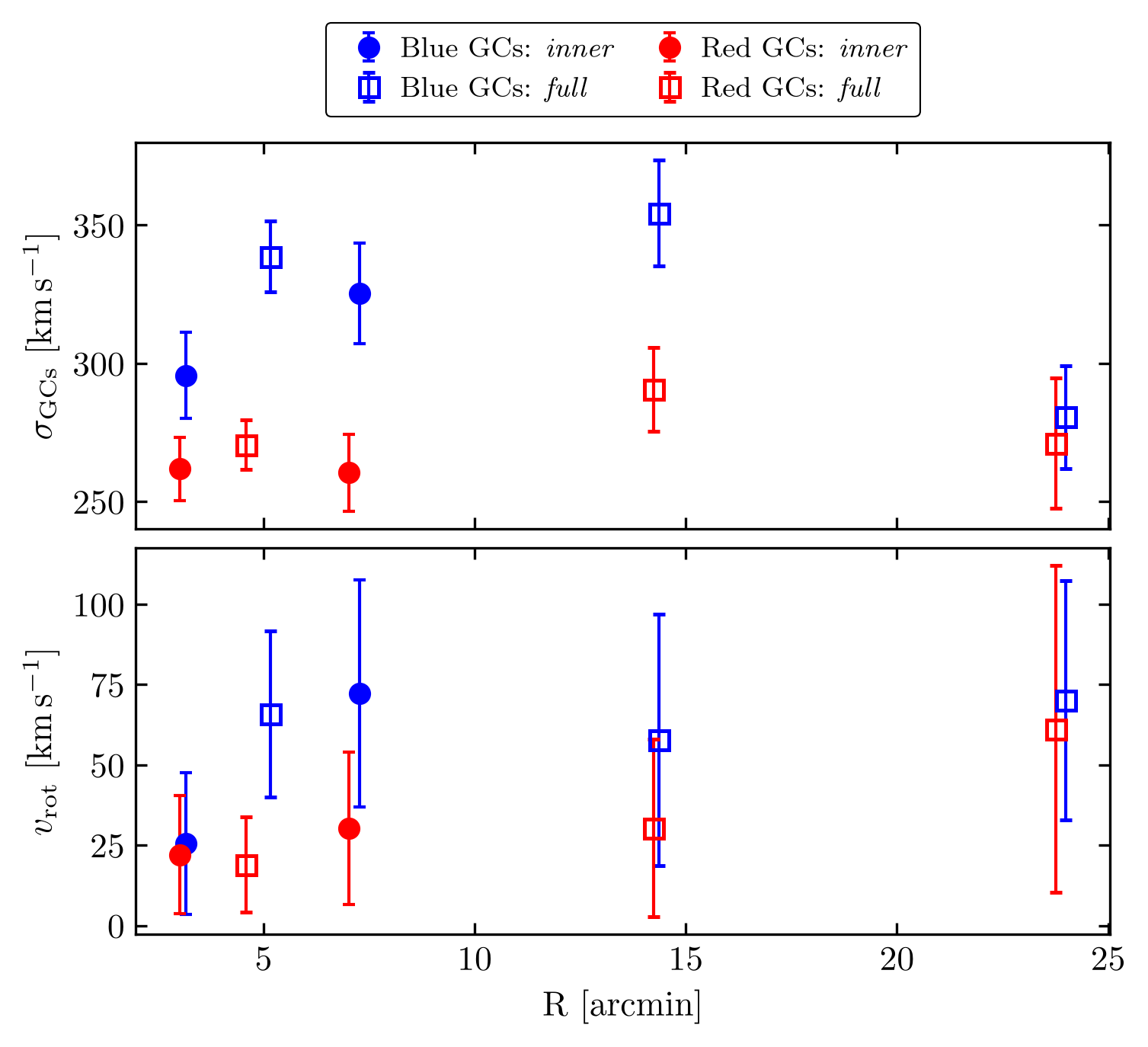}
\caption{GC rotational analysis showing the velocity dispersion (top panel) and rotational amplitude (bottom panel) of the red and blue GCs in the \textit{inner} and \textit{full} samples. In both panels, the filled red and blue circles show the measurements for the red and blue GCs in the \textit{inner} sample, respectively, while the open squares show the corresponding measurements for the \textit{full} sample.}
\label{fig:vrot_GCs}
\end{figure}

\subsection{GCs rotation}

To check for any rotational signature in the GC population, we modelled the GC kinematics as a function of position angle with a simple model that describes the rotational amplitude and velocity dispersion. We adopted a fitting methodology similar to the approach of \cite{Katja2020} (for details see their section 4.2.1) first applied in \cite{Cote2001} (see also \citealt{napolitano2001} for a test of the effectiveness of the method).

Figure \ref{fig:vrot_GCs}, shows the modelled GCs kinematics illustrating the rotational amplitude ($v_{rot}$) and velocity dispersion ($\sigma_{GCs}$). We first start by analysing the \textit{inner} sample. We divided the GCs into two radial bins, each 5 arcminutes wide. The red GCs show no significant rotational signature, with a rotational amplitude below 30 \kms. In contrast, the blue GCs exhibit a rotational amplitude of $\sim74 \pm35$ \kms and an angle of the rotation axis of 126 degrees. These values agree with the previous rotational analysis of NGC~1399 GCs by \citet[][see sections 7 and 7.4]{Schuberth2010}. However, we find a lower rotational amplitude for the blue GCs compared to \cite{Schuberth2010}, which may be attributed to differences in the GC samples used.

For the \textit{full} sample, we divided the sample into three bins, each 10 arcminutes wide. Similar to the \textit{inner} sample, we do not see any rotation signature for the red GCs. Only in the outer bins between 20 to 30 arcminutes, we find a rotational amplitude of $\sim60\pm50$ \kms, though this is not significant due to the large uncertainty. For the blue GCs, we observe a mild rotation with an amplitude of $\sim65\pm25$ \kms within 20 arcmin. In terms of $v_{rot}$/$\sigma$, the blue GCs have a value of $\sim$0.22, showing a slowly rotating system. Given these low rotational signature for both red and blue GCs population, we can safely consider the assumption of no rotation in our dynamical modelling. 

% FIGURE 6
\begin{figure*}[h]
\sidecaption
    \centering
    \includegraphics[width=19cm]{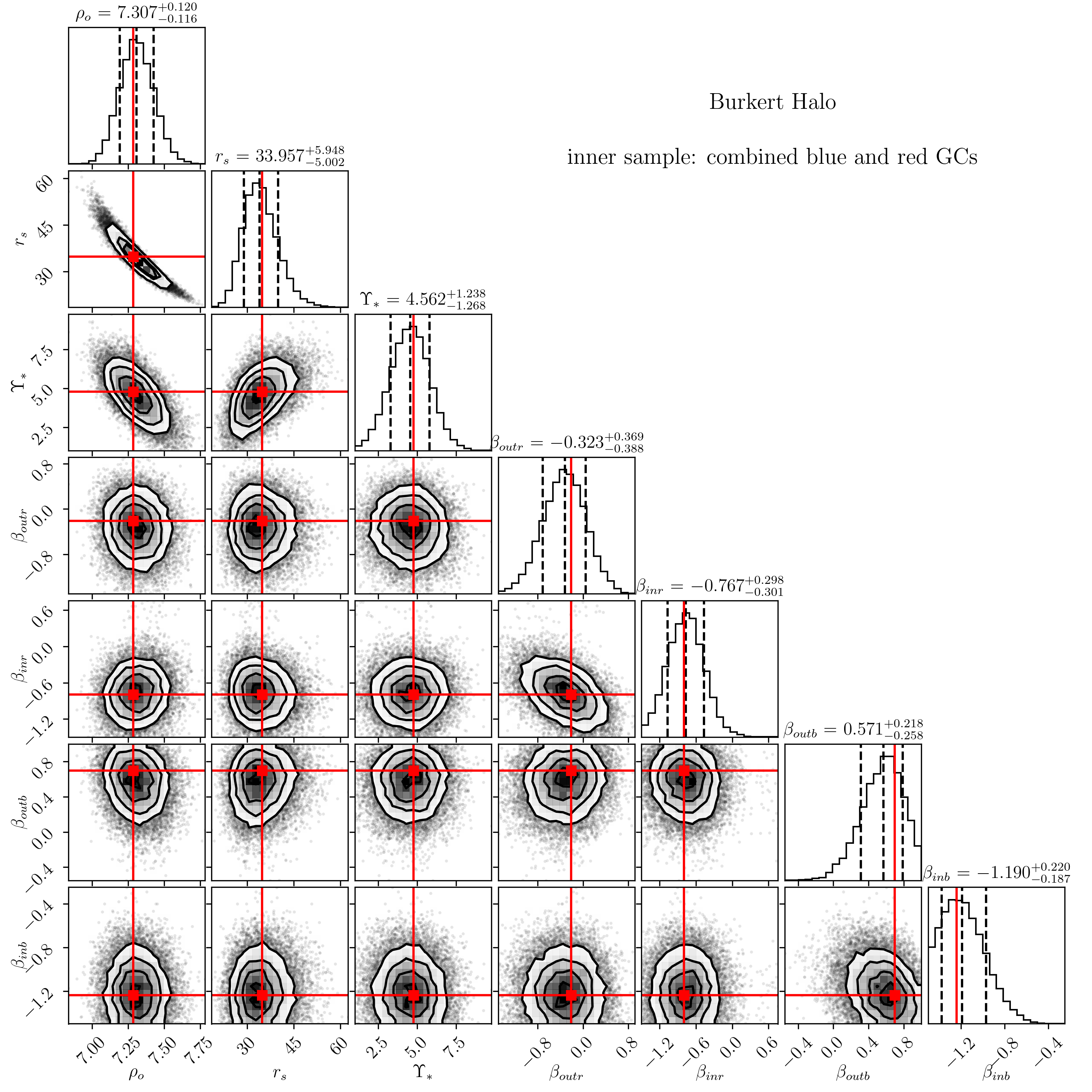}
    \caption{1D and 2D posterior distribution of parameters (MCMC output) from the two-component Jeans modelling using the Burkert halo for the
\textit{inner} sample. The histogram on the top shows the 1D distribution of the parameters, and error bars mark the $\pm$16 and 84 percentiles.
In the 2D projection, the ellipse denotes the 1, 2, 3$\sigma$ region of the projected covariance.}
    \label{fig:bur_NGC~1399_corner}
\end{figure*}

% FIGURE 8
\begin{figure*}[h]
    \centering
    \sidecaption
    \includegraphics[width=18cm]{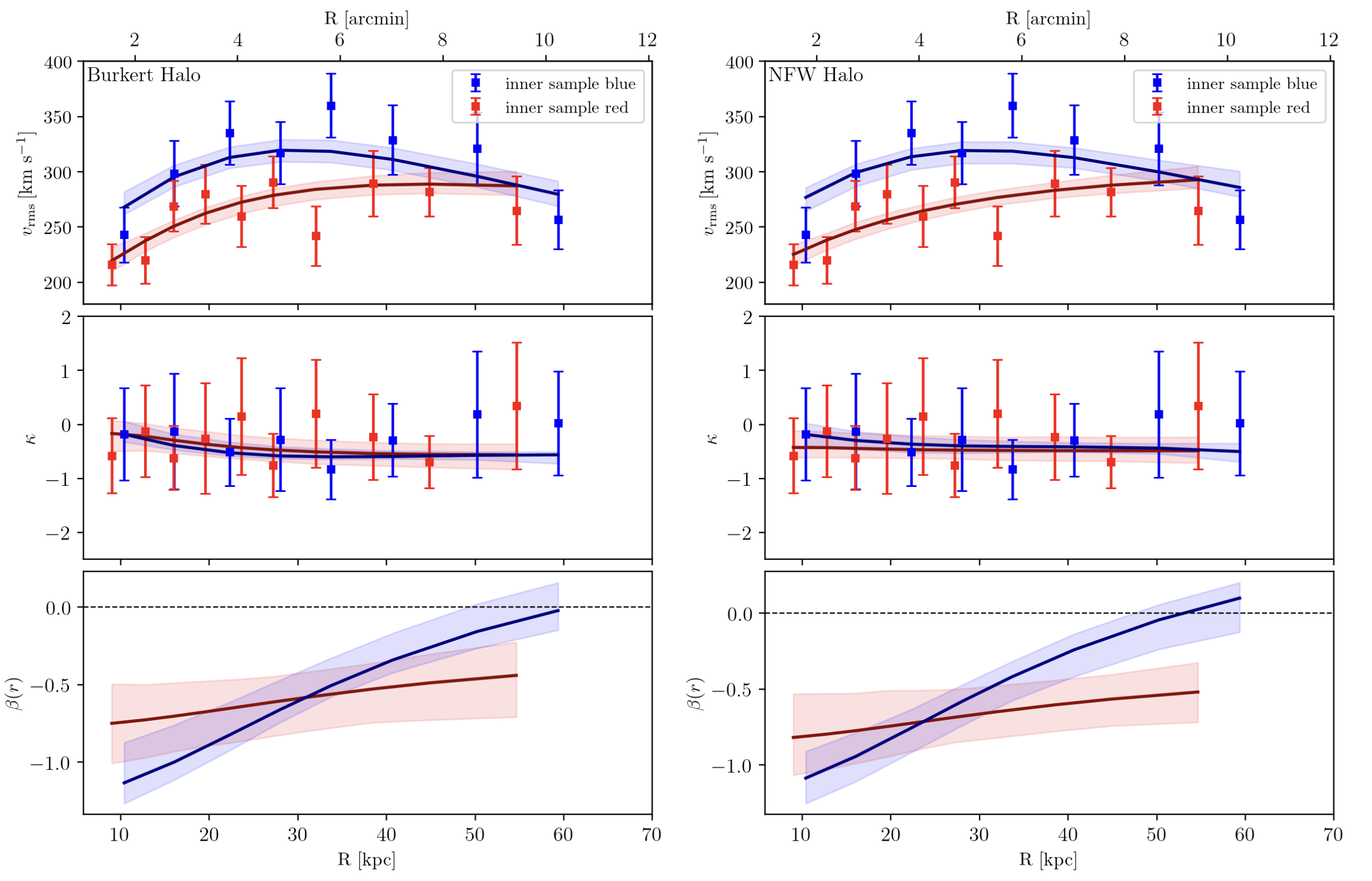}
    \caption{Observed and modelled velocity dispersion and kurtosis profiles of the \textit{inner} sample, as defined in Section \ref{sec2}. Left: Best-fit model for the Burkert halo. The top and middle panels display the velocity dispersion and kurtosis profiles for the red and blue GCs, with the solid line representing the best-fit model. The lower panel shows the anisotropy profile derived from the best-fit parameters (Table \ref{Tab:best_fit_params}) using Equation \ref{equ9}. Right: Same as the left panel but for the NFW halo best-fit parameters.}
    \label{fig:best_fit_param_NGC~1399}
\end{figure*}

%### FIGURE 11 ###
\begin{figure*}[h]
    \centering
    \includegraphics[width=18cm]{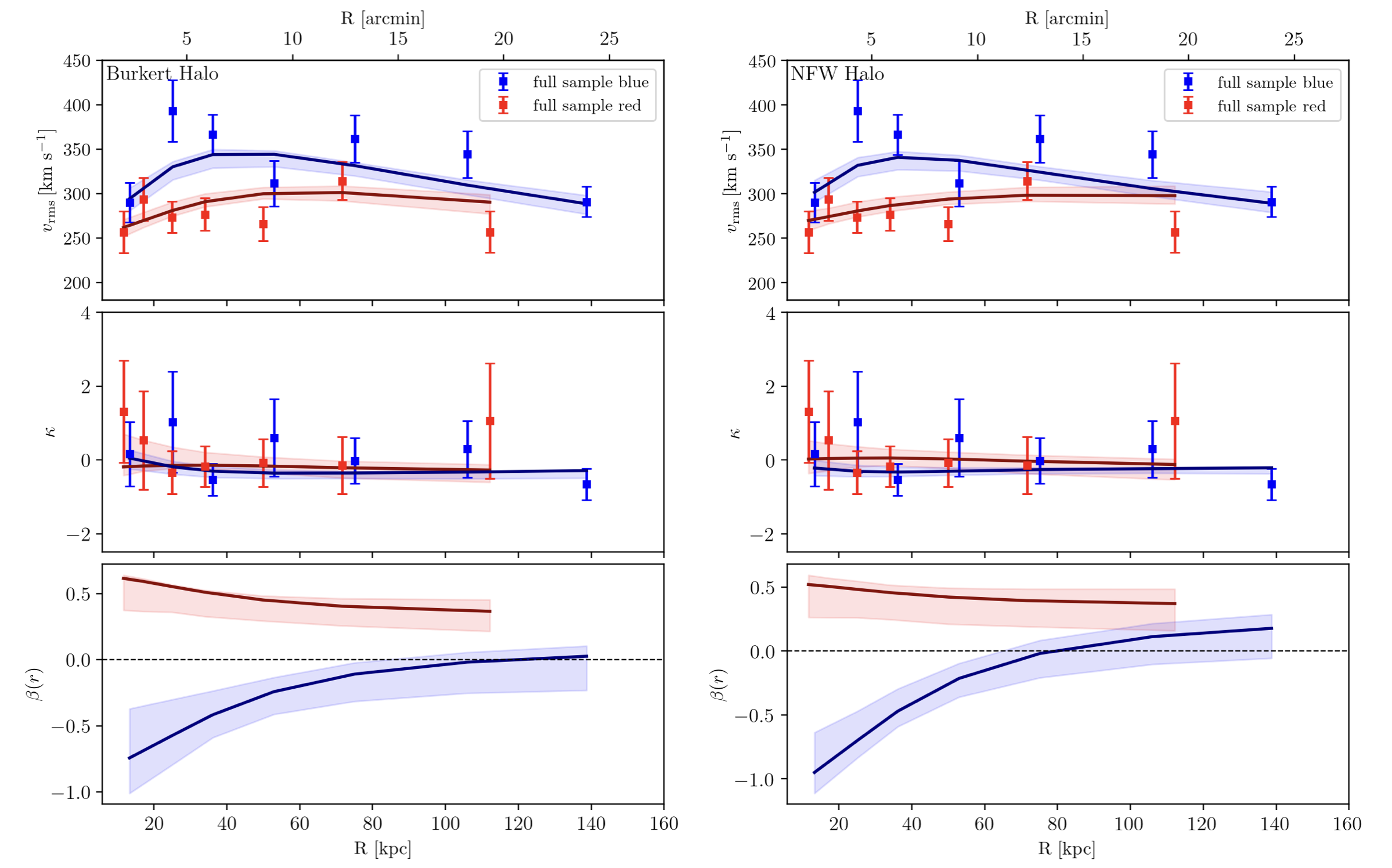}
    \caption{Observed and modelled velocity velocity dispersion and kurtosis profiles of \textit{full} sample. Left: Best fit model for Burkert halo. Top and middle panels show the velocity dispersion and kurtosis profiles for the red and blue GCs (squares) and solid line indicates the best fit model. Right: Same as left but for the NFW halo best parameters.}
    \label{fig:best_fit_param_NGC~1399+IC}
\end{figure*}

% ########################### Section 3 ##########
\section{Dynamical modelling}\label{sec3}

We use the GCs tracer density (Sect. \ref{sec2:tracerdensity}) and kinematic information to obtain the baryonic and DM mass profile of NGC~1399. We apply higher order Jeans moments for the dynamical modelling work, similar to what was used in \cite{napolitano2009, napolitano2011}. For the specific application to the modelling of red and blue GCs, see \citet{napolitano2014}.  Here, we briefly summarise the Jeans modelling framework and describe our methodology.

\subsection{Jeans modelling framework}

We model the dynamics of NGC~1399 using spherical Jeans equations under the assumptions of dynamical equilibrium and negligible rotation. The tracer density profiles of red and blue GCs, together with the total gravitational potential, are used to predict the line-of-sight velocity dispersion and kurtosis profiles. To reduce the mass--anisotropy degeneracy, we jointly model the second and fourth velocity moments of the LOSVD. The orbital structure of the tracers is described through a parametric velocity anisotropy profile, $\beta(r)$ introduced by \cite{churazov2010}, characterised by inner and outer anisotropy parameters ($\beta_{in}$, $\beta_{out}$) and a transition radius $r_a$. The modelled gravitational potential in our dynamical modelling consist of two components stellar and DM halo mass. For the DM halo we have used both a core and cupsy DM halo profile parametrised as Burkert \citep{Burkert1995} and NFW profile \citep{Navarro1997}. Full details of the Jeans equations, projection formalism, anisotropy parametrisation and gravitational potential are given in Appendix \ref{Jeans_model}. In the following subsection, we describe the procedure used to estimate the best-fit model parameters.

\subsection{Maximum likelihood analysis and parameter inference}

We employ a maximum likelihood (ML) approach to infer the best-fit parameters used in our dynamical modelling. Differently from \citet{napolitano2014}, who used a simpler $\chi^2$ minimization on a regular grid of parameters, here we have implemented a Bayesian analysis. In this approach, the posterior probability distribution is the product of the likelihood function of the observation (the data in our case are velocity dispersion or kurtosis) given the model with some parameters and priors on the model parameters. The set of model parameters that maximize the likelihood function produces a model that best fits the data. The total likelihood is the product of the individual measurements of the dataset. 

In our case, we create two likelihood functions corresponding to the velocity dispersion and kurtosis of the GCs. Assuming a Gaussian LOSVD, for a given population $k$ of GCs (either red or blue) with a standard deviation equal to the uncertainty of the measured $v_{rms}$, the log-likelihood of observing a GC population having a velocity dispersion $v_{rms}$, given a set of model parameters $\textbf{\textit{M}} = \bigl\{r_{s}, \rho_{o}, \Upsilon_{*}, \beta_{in}, \beta_{out} \bigl\}$ is:

\begin{equation}
    \ln  {p_{v_{rms}}} ({v_{rms_{k}}|\textbf{\textit{M}}_{k}}) =
    -\frac{1}{2} \sum_{i=1}^{N} \left[ \left(\frac{v_{rms,i} - \sigma_{k}}{\Delta v_{rms,k}} \right)^{2}  + \ln (2\pi\Delta v_{rms,i}^{2})\right]
\end{equation}

Here $N$ is the number of bins used in the $v_{rms}$ profile and $k$ denotes either red or blue GCs population. Similarly, we have another log-likelihood term for kurtosis ($\kappa$), and the combined log-likelihood is written as:

\begin{equation}
    \ln L = \ln p_{v_{rms}} + \ln p_{\kappa}
\end{equation}

Finally, when modelling red and blue GCs kinematics together, we get the total log-likelihood expressed as :
\begin{equation}
    \mathcal{L} = \sum_{k=1}^{2} \ln L_{k}
\end{equation}

We performed a joint two component tracer modelling i.e. fitting both red and blue GCs kinematics together with each having a separate log-likelihood function. This allows to provide a stronger constraint on the orbital distribution of red and blue GCs than fitting each tracer individually (either red or blue). For this setup, we have a total of seven input parameters including the scale radius and density normalization parameter of the DM density profile, stellar mass-to-light ratio, and two anisotropy terms corresponding to each red and blue GCs as: $\bigl\{ r_{s}, \rho_{o}, \Upsilon_{*}, \beta_{inr}, \beta_{outr}, \beta_{inb}, \beta_{outb} \bigl\}$.

To explore the parameter space and to find the best-fit parameters, we utilize the EMCEE package \citep{emcee2013}, a python-based implementation of the affine-invariant Markov Chain Monte Carlo (MCMC) sampling algorithm. The EMCEE package uses several walkers to explore the parameters space and uses a specified number of steps to converge to best-fit values. 

For the prior distributions, we used both a flat (uniform) and Gaussian prior. A flat prior assumes that all values within a defined range are equally probable, representing a non-informative assumption. In contrast, a Gaussian prior assumes a normal distribution centred around a specific mean with a standard deviation, thereby favouring the values near the mean value and suppressing the values farther from it. The use of the Gaussian prior is motivated by the fact that it helps to account for degeneracy among the parameters, especially between the $r_{s}$ and $ \rho_{o}$ of the DM halo profile. For both uniform and Gaussian priors, we used 100 walkers and ran the MCMC for 10,000 steps to ensure the parameter space was well explored and converged. For uniform priors we varied each within the following range: $r_{s}: \bigl\{10,300\bigl\}$ kpc, log $\rho_{o}: \bigl\{ 5.5, 9.5\bigl\}$ $M_{\odot}/{\rm kpc}^{3}$, $\Upsilon_{*}: \bigl\{1,10\bigl\}$, and  $\beta_{in} = \beta_{out}: \bigl\{-1.5,1.0\bigl\}$. To construct the Gaussian priors, we used the mean and standard deviation of each parameter obtained from the runs with the flats prior as the Gaussian parameters (mean and standard deviation). Our initial tests showed that, with the uniform flat prior, the autocorrelation time of the MCMC chains was very large, indicating slow exploration of the parameter space. In contrast to this, the Gaussian priors led to significantly improved sampling efficiency and the parameters were well converged and constrained. Although the best-fit values obtained from the flat and Gaussian priors are broadly consistent, we adopt the results from the Gaussian-prior runs as our final best-fit parameter set due to their improved convergence and stability.

% ################### SECTION 4 #################################

\section{Results}\label{sec4}

This section presents the results from our dynamical modelling and fit to the GCs kinematics and best-fit parameters.

% TABLE 2 ##############
%### Best fit parameters table ###
\begin{table*}
\caption{Best fit parameter of the two-component joint modelling.}
\label{Tab:best_fit_params}
\centering
\setlength{\tabcolsep}{3pt}
\renewcommand{\arraystretch}{1.6} % Default value: 1
  \begin{tabular}{c|lllllllll|c}
    \toprule
    \multicolumn{1}{c|}{\textbf{Sample}} &
      \multicolumn{9}{c|}{} &
      \multicolumn{1}{c}{} \\
      \hline
      \multicolumn{1}{c|}{\textbf{\textit{inner}}} &
      \multicolumn{9}{c|}{Parameters} &
      \multicolumn{1}{c}{} \\
      \hline
    
       & $\rho_{o}$ & $r_{s}$ & $\Upsilon_{*}$ & $\beta_{inr}$ & 
       $\beta_{outr}$ & $\beta_{inb}$ & $\beta_{outb}$ & log $M_{200}$ & $R_{200}$ & $\mathcal{L}$  \\
      & $ [M_{\odot}/kpc^{3}] $ & $[kpc]$ & $ [M_{\odot}/L_{\odot, g}]$ & & & & & $M_{\odot}$ & kpc & \\

    \midrule
   
Burkert  & $7.39^{+0.12}_{-0.11}$ & $31.01^{+5.77}_{-4.94}$ & $4.04^{+1.30}_{-1.25}$ & $-0.77^{+0.29}_{-0.30}$ & $-0.26^{+0.38}_{-0.38}$ & $-1.24^{+0.22}_{-0.18}$ & $0.53^{+0.21}_{-0.26}$  & $13.29^{+0.07}_{-0.08}$ & $569.58^{+33.47}_{-37.59}$ & -95.57 \\

NFW  & $6.29^{+0.18}_{-0.17}$ & $123.48^{+43.43}_{-32.80}$ & $3.35^{+1.11}_{-1.02}$ & $-0.84^{+0.27}_{-0.27}$ & $-0.34^{+0.34}_{-0.34}$ & $-1.20^{+0.20}_{-0.17}$ & $0.69^{+0.20}_{-0.23}$  & $13.75^{+0.15}_{-0.13}$ & $812.72^{+101.93}_{-79.73}$ & -96.17 \\

    \midrule
      \multicolumn{1}{c}{\textbf{\textit{full}}} &
      \multicolumn{9}{c}{} &
      \multicolumn{1}{c}{} \\
      \hline
            
Burkert & $7.15^{+0.09}_{-0.09}$ & $47.12^{+5.90}_{-5.90}$ & $6.08^{+1.76}_{-1.88}$ & 
$0.63^{+0.11}_{-0.15}$ & $0.33^{+0.15}_{-0.12}$ & $-0.83^{+0.34}_{-0.32}$ & $0.09^{+0.19}_{-0.20}$  & $13.49^{+0.06}_{-0.05}$ & $662.34^{+33.79}_{-26.08}$ & -77.60  \\

NFW  & $6.26^{+0.16}_{-0.15}$ & $135.91^{+33.80}_{-27.96}$ & $4.59^{+1.85}_{-1.87}$ &
$0.53^{+0.15}_{-0.23}$ & $0.34^{+0.14}_{-0.22}$ &  $-1.08^{+0.29}_{-0.28}$ & $0.28^{+0.18}_{-0.20}$ & $13.81^{+0.09}_{-0.09}$ & $846.71^{+61.63}_{-61.15}$ & -78.02 \\

    \bottomrule
  \end{tabular}
\end{table*}

\subsection{Model fit to the \textit{inner} sample}

We begin with the joint-component modelling of the \textit{inner} sample. Figure \ref{fig:bur_NGC~1399_corner} shows the 1D and 2D posterior distributions of the parameters for the Burkert halo. Each parameter has a smooth unimodal distribution suggesting that parameters best-fit values and uncertainties are reasonably well constrained. A very similar behaviour is found for the NFW halo, as shown in Figure \ref{fig:nfw_NGC~1399_corner}.

The scale radius ($r_{s}$) and density normalisation ($\rho_{s}$) show a strong degeneracy. This is expected, as different combinations of these parameters can produce a similar enclosed mass profile. In contrast to the Burkert halo, the scale radius of the NFW halo has is large with a large uncertainty. The resulting stellar mass-to-light ratio ($\Upsilon_{*}$) is slightly higher for the Burkert halo, in comparison to the NFW halo, but they both agree within the errors. For both halos, the observed orbital anisotropy for both, red and blue GCs, is very similar. The red GCs are tangentially anisotropic within 10 arcmin, while the blue GCs are tangentially anisotropic within the inner 10 arcmin and gradually become isotropic at larger radii. In the NFW case, the blue GCs show a mild radial anisotropy in the outskirts.

Table \ref{Tab:best_fit_params} summarises the best-fit values corresponding to the maximum posteriori solution of the parameters for both models. Figure \ref{fig:best_fit_param_NGC~1399} shows the modelled GCs velocity dispersion and kurtosis profile for both Burkert and NFW halos for the \textit{inner} sample. Examining the best-fit models, it is evident that both models can re-produce the observed kinematics quite well, making it difficult to distinguish a preferred model. 

The total log-likelihood values of the Burkert and NFW models are very similar, with the Burkert halo giving a slightly higher likelihood ($\ln \mathcal{L} = -95.57$) than the NFW halo ($\ln \mathcal{L} = -96.17$). Thus, the Burkert halo provides a marginally better fit to the \textit{inner} sample. Nevertheless, the difference is small, indicating that the present data do not strongly distinguish between the two halo models. To assess these two models in the context of the $\Lambda$CDM predictions, we have calculated the virial mass $M_{200}$ and virial radius $r_{200}$. Here $M_{200}$ is defined as the enclosed mass within $r_{200}$, such that the average density within $r_{200}$ is 200 times the critical density of the Universe. Table \ref{Tab:best_fit_params} lists the corresponding $M_{200}$ and $r_{200}$ values for both models. For the \textit{inner} sample, the inferred virial masses differ substantially between the two halo profiles. We obtain $\log M_{200} = 13.29^{+0.07}{-0.08}$ for the Burkert halo and $\log M{200} = 13.75^{+0.15}_{-0.13}$ for the NFW halo. Thus, the NFW halo favours a virial mass higher by $\sim 0.46$ dex. This indicates that, although the two halo models provide similarly good fits to the kinematic data, they imply significantly different halo masses for the \textit{inner} sample.

\subsection{Model fit to \textit{full} sample}

Similar to the \textit{inner} sample, we found that both the Burkert and NFW halos, reproduce the observed kinematics well for the \textit{full} sample. Figures \ref{fig:bur_NGC~1399+IC_corner} and \ref{fig:nfw_NGC~1399+IC_corner} show the 1D and 2D posteriors distributions of the fit parameters, respectively. As with the \textit{inner} sample, the model parameters are well constrained for both halos, with uncertainties properly determined. In contrast to the \textit{inner} sample, the uncertainty on the scale radius for the NFW halo is significantly lower and well constrained. The stellar mass-to-light ratio $\Upsilon_{*}$, obtained for the \textit{full} sample is higher than that for the \textit{inner} sample. 

Figure \ref{fig:best_fit_param_NGC~1399+IC} shows the best-fit models to the observed kinematics of the \textit{full} sample. Compared to the \textit{inner} sample, we find a different anisotropy behaviour for the red GCs in the inner region. For both halos, the red GCs exhibit radial anisotropy within 60 kpc and tend toward isotropy beyond this radius. In the case of the blue GCs, we find that within 40 kpc, they show strong tangential anisotropy, transitioning to mild tangential anisotropy within 80 kpc. Beyond this, the blue GCs exhibit an isotropic or mildly radial behaviour with increasing radius. As in the \textit{inner} sample, the total log-likelihood values for the \textit{full} sample are very similar for the two halo models, with the Burkert halo yielding a slightly higher value. Thus, the Burkert halo provides a marginally better fit, although the difference remains small. The inferred virial masses also differ between the two halo profiles, with $\log M_{200} = 13.49^{+0.06}_{-0.05}$ for the Burkert halo and $\log M_{200} = 13.81^{+0.09}_{-0.09}$ for the NFW halo. Therefore, the NFW halo favours a higher virial mass, although the offset is less extreme than that found for the \textit{inner} sample.

\subsection{Comparison of the stellar mass-to-light ratio with previous studies}\label{sec4.3}

The best-fit stellar mass-to-light ratios in the \(g\) band in our models are \(\Upsilon_{*,g} = 4.04^{+1.30}_{-1.25}\) and \(3.35^{+1.11}_{-1.02}\) for the \textit{inner} sample, and \(\Upsilon_{*,g} = 6.08^{+1.76}_{-1.88}\) and \(4.59^{+1.85}_{-1.87}\) for the \textit{full} sample, for the Burkert and NFW haloes, respectively. Our best-fitting values are lower than the estimate of \citet{Saglia2000}, who found \(M/L_{B} \sim 10\) for the luminous component of NGC~1399. However, this comparison is not direct because of the different photometric band, radial coverage, and modelling assumptions. \citet{Richtler2004} converted the \(B\)-band value derived by \citet{Saglia2000} and obtained \(M/L_{R} = 5.5\), which is close to the value reported by \citet{Gebhardt2007}, who found a stellar \(M/L_{R} = 5.77 \pm 0.4\). To enable a comparison with the \(R\)-band values reported in previous studies, we approximately converted our \(g\)-band mass-to-light ratios to the \(R\) band. This gives \(\Upsilon_{*,R} = 3.39^{+1.09}_{-1.05}\) and \(2.81^{+0.93}_{-0.86}\) for the \textit{inner} sample, and \(\Upsilon_{*,R} = 5.11^{+1.48}_{-1.58}\) and \(3.86^{+1.55}_{-1.57}\) for the \textit{full} sample, for the Burkert and NFW haloes, respectively. These converted values are broadly comparable to the \(R\)-band estimates reported in the literature. In particular, the \textit{full}-sample Burkert value is in close agreement with the \(M/L_{R} \sim 5.5\) estimate of \citet{Richtler2004}, while the other values are somewhat lower but remain of the same order.

Our values are also broadly consistent with the stellar-population analysis of NGC~1399 by \citet{Vaughan2018}. They found a central stellar-population value of \(M/L_{V} = 9.0^{+1.69}_{-1.44}\), decreasing to \(5.48^{+4.43}_{-2.27}\) beyond one effective radius. For the \textit{full} sample our best-fit mass-to-light ration in V band is $5.43^{+1.57}_{-1.68}$ and $4.10^{+1.65}_{-1.67}$ for the Burkert and NFW halo, respectively, which is in close agreement with the results of \cite{Vaughan2018}. Therefore, our inferred values are lower than the central estimates reported in the literature, but remain broadly comparable to the values of \(M/L \sim 5\)--6 reported or adopted at larger radii by \citet{Gebhardt2007}, \citet{Schuberth2010}, and \citet{Vaughan2018}.

\section{Discussion}\label{sec5}
In this section, we discuss and compare our modelling results and the orbital anisotropy of GCs to previous studies of NGC~1399 and simulations studies within the $\Lambda$CDM framework. 

\begin{figure}[htp]
    \centering
    \includegraphics[width=1.05\linewidth]{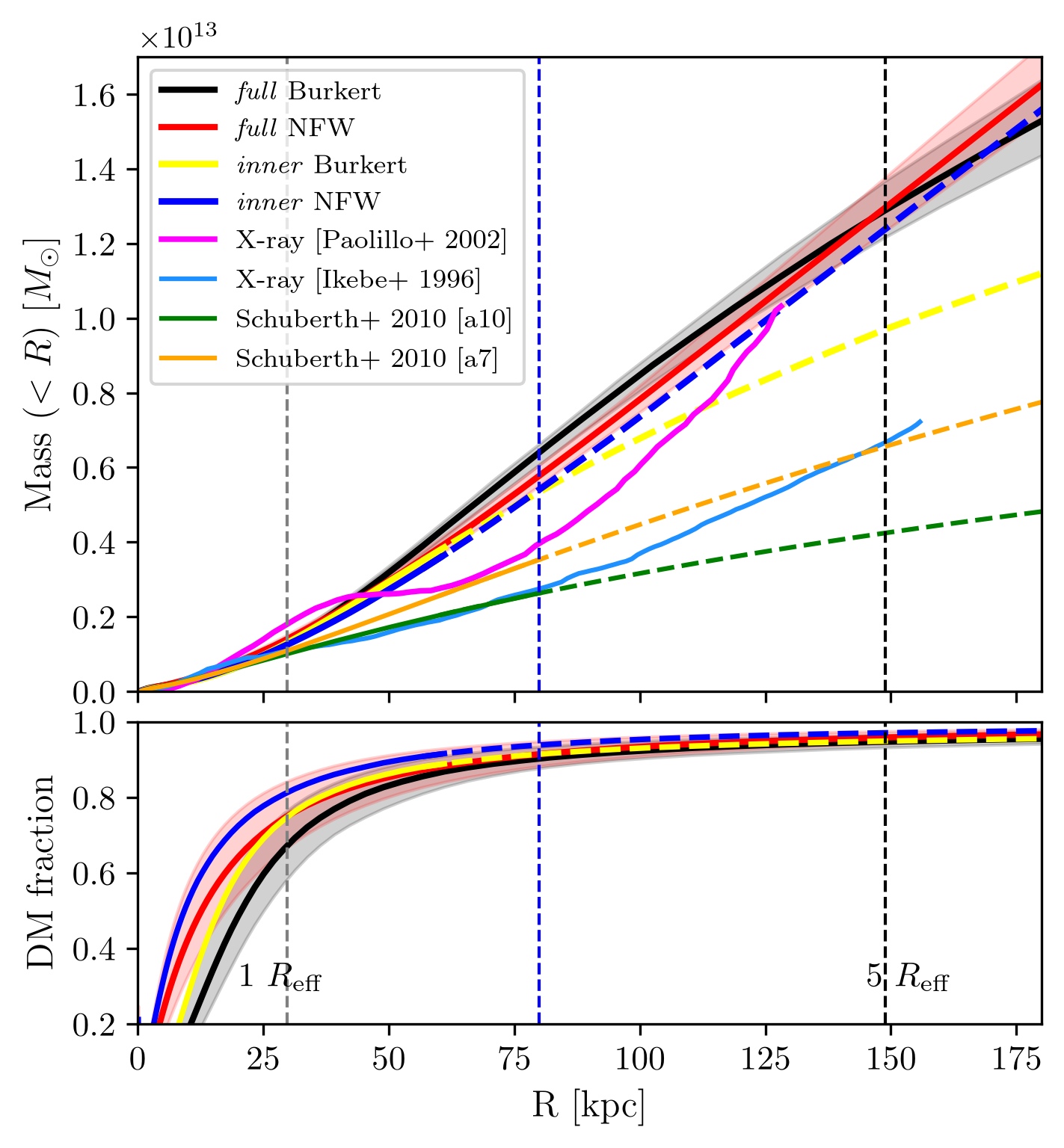}
    \caption{Enclosed mass profile of NGC~1399. The black and red solid lines indicate the enclosed mass for the \textit{full} sample assuming Burkert and NFW halos, respectively. The solid yellow and blue lines indicates the mass profile for the \textit{inner} sample for Burkert and NFW halos, respectively, while the dashed lines show the extrapolations of these mass profiles. Vertical dashed grey and black lines mark the 1$\rm Re$ and 5$\rm Re$ radii, respectively. The vertical blue dashed line indicates the radial extent of the GC sample used in \cite{Schuberth2010} to derive the mass profile of NGC~1399. The solid orange and green curves represent the mass profiles of NGC~1399 obtained from the 'a10' and 'a7' models of \cite{Schuberth2010}, while the dashed orange and green curves indicate their extrapolations. Bottom panel: DM fraction profile inferred from the \textit{inner} and \textit{full} sample for the Burkert and NFW halos, using the same colours and line styles as in the top panel.}
    \label{fig12:mass_profile}
\end{figure}

\subsection{Mass profile}\label{sec5.1}

NGC~1399 has been extensively studied using various kinematical tracers, allowing us to compare our mass estimates to previous studies in the literature. The earliest attempts to dynamically understand NGC~1399 were made by \cite{Saglia2000, Napolitano2002, Richtler2008, Schuberth2010}. Figure \ref{fig12:mass_profile} shows the total enclosed mass profile obtained from our best-fit models and comparison with previous works. The solid black and red curves indicate the mass profiles for the Burkert and NFW profiles for the \textit{full} sample, respectively, whereas the dashed curves correspond to the \textit{inner} sample. For clarity, we show the 1 $\sigma$ uncertainty only for the mass profiles of the \textit{full} sample. 

We compare the \textit{inner} and \textit{full} mass profiles to the extent of the available GC kinematics, i.e., $\sim$10 arcminutes (60 kpc) for the \textit{inner} sample and $\sim$25 arcminutes (150 kpc) for the \textit{full} sample. For the \textit{inner} sample, we find that within 60 kpc, the mass profiles for both the Burkert (yellow solid line) and NFW (blue solid line) halos agree well. The dashed yellow and blue lines indicate the extrapolated \textit{inner} mass profiles of the Burkert and NFW halos, respectively. These profiles begin to diverge beyond 75 kpc, with the discrepancy becoming more apparent at $\sim$150 kpc. In contrast, the mass profiles derived from the \textit{full} GC sample which include ICGCs show excellent agreement between the NFW and Burkert halos.

The bottom panel of Figure \ref{fig12:mass_profile} shows the dark matter (DM) fraction for both Burkert and NFW halos, using the inner and full samples and following the same line styles as in the top panel. At 1 $\rm Re$, the DM fraction in the inner sample exceeds 50\% for both halos—reaching approximately 70\% for the Burkert profile and 80\% for the NFW profile. In contrast, the full sample yields slightly lower DM fractions: around 60\% for the Burkert halo and 70\% for the NFW halo. At 5 $\rm Re$, the DM fraction exceeds 90\% in all cases. 

Our measured DM fractions at 1 $\rm Re$ are notably higher than typical values reported in the literature for early-type galaxies, which often lie in the range of $\sim$ 10–20 \% when derived from stellar or discrete tracers such as GCs \citep{Cappellari2013DM, Zhu2016, Dumont2024}. However, there are also examples of systems with similarly high DM fractions within 1 $\rm Re$. For instance, \cite{Chao2020} reported a DM fraction of $\sim$73\% for M87 at 1$\rm Re$, increasing to $\sim$94\% at 5$\rm Re$, based on GC dynamical modelling extending out to 400 kpc. One caveat of our analysis is the assumption of a constant stellar mass-to-light ratio in the dynamical modelling. Allowing for a radially varying mass-to-light ratio could influence the inferred DM fraction. Nonetheless, our results strongly suggest that NGC 1399 is a dark matter-dominated galaxy, with the DM fraction exceeding 50\% beyond 1$\rm Re$.

Previous mass-profile measurements of NGC~1399 using GCs data have been performed by \cite{Richtler2008, Schuberth2010, Samurovic2016}. Here, we primarily compare our results with \cite{Schuberth2010}. They used spherical Jeans modelling and GC kinematics within 80 kpc to derive the mass profile of NGC~1399. Additionally, they incorporated stellar kinematics within the inner 10 kpc \citep{Saglia2000}. By assuming tangential, radial and isotropic anisotropic profiles, they performed various sets of models. The mass profile from their best-fit model, 'a10', which incorporates red GCs and stellar kinematics (for details, see their Section 10), is shown as an orange curve in Figure \ref{fig12:mass_profile}. Another mass profile from their model, 'a7', which includes both red and blue GCs, is shown in light green. The dashed orange and green curves show the extrapolations of these mass profiles. We find that within 1$\rm Re$ our mass profiles from both samples are consistent with the 'a7' and 'a10' model of \cite{Schuberth2010}. However after 1$\rm Re$ our mass profiles show a higher enclosed mass. 

The difference between our enclosed mass profile and that of \cite{Schuberth2010} may have arise due to several factors. One of the main reasons could be the use of different GC samples. When comparing the velocity dispersion of our \text{inner} sample, which extends to 60 kpc similar to the physical scale of the GC sample used by \cite{Schuberth2010}, we found that, although the blue GCs have similar velocity dispersion values to those in their work, the red GCs in our sample have values approximately 50–80 \kms higher. In the case of the full sample, which extends to 150 kpc, the velocity dispersion profiles of both red and blue GCs were higher than those reported by \cite{Schuberth2010}. The inclusion of ICGCs resulted in a flattened dispersion profile out to 150 kpc, thereby leading to a higher enclosed mass profile. Additionally, we employed a more generalized modelling approach by fitting both the velocity dispersion and kurtosis profiles. Our model also incorporated a more flexible anisotropy profile, rather than assuming purely radial or tangential anisotropy.

The mass profile of NGC~1399 and the Fornax Cluster, as derived from X-ray studies, has been previously presented by \cite{Ikebe1996} and \cite{Paolillo2002}. \cite{Ikebe1996} first reported the presence of substructures in the X-ray emission of the Fornax Cluster. Later, using ROSAT observations, \cite{Paolillo2002} confirmed these findings and identified three distinct substructures associated with the central galaxy NGC~1399, its extended halo, and the cluster environment. In Figure \ref{fig12:mass_profile}, the light blue and magenta curves represent the X-ray mass profiles from \cite{Ikebe1996} and \cite{Paolillo2002}, respectively. Within the inner 20 kpc (approximately 1 $\rm Re$), our mass profiles—derived from both the \textit{inner} and \textit{full} GC samples are in good agreement with the X-ray estimates. However, beyond 1 $\rm Re$, the X-ray mass profiles consistently lie below our GC-based estimates.

The X-ray mass profile from \cite{Paolillo2002} exhibits shoulder-like features near the interfaces of different X-ray regions, indicating the presence of substructures. These features complicate a direct comparison with our mass profiles, which are derived from the kinematics of GCs and trace the global gravitational potential more uniformly. Around $\sim$110 kpc, the X-ray profile shows good agreement with our full sample mass profile. Overall, our full sample profile broadly agrees with the mass profile presented by \cite{Paolillo2002}, despite the localized deviations introduced by X-ray substructure.

One of the limitations of our dynamical modelling is that we considered only cuspy (NFW) and cored (Burkert) DM profiles, rather than allowing for a generalized inner slope of the DM halo. Several studies in the literature suggest that dynamical friction, active galactic nucleus (AGN) feedback, and baryonic processes can alter the inner slope of the DM profile \citep{Blumenthal1986, Schaller2015, Peirani2017}. The inner DM profile is influenced by the stellar distribution in galaxies, making it crucial to consider a generalized DM halo profile \citep{Sand2004, Sand2008, Newman2013}. As mentioned earlier, in a future study, we plan to investigate the DM halo distribution using a generalized NFW profile, which will provide tighter constraints on the DM distribution. However, given that we have already explored both core and cuspy DM halos, we speculate that even with a generalized NFW profile, our total enclosed mass profile will fall between those obtained using the current Burkert and NFW models for the \textit{full} sample.

In summary, we find that regardless of the DM halo profile used, the mass profiles for the \textit{full} and \textit{inner} samples agree within 1$\sigma$ uncertainty. We found that for both \textit{inner} and \textit{full} sample of GCs, we measure an enclosed mass profile higher than the previous estimate of \cite{Schuberth2010}, but consistent with the enclosed mass profile obtained from X-ray observations \citep{Paolillo2002}. For the \textit{inner} sample, the mass profiles derived from NFW and Burkert halos begin to diverge beyond $\sim$75 kpc. In contrast, the profiles for the \textit{full} sample which includes ICGCs remain consistent across both halo models, indicating that the inclusion of ICGCs reduces the DM halo profile dependence.

The inclusion of ICGCs in the \textit{full} sample results in a flattening of the velocity dispersion profile, which potentially leads to a higher enclosed mass profile and suggests that, in addition to measuring the NGC~1399 DM fraction, we are also measuring the the gravitational potential of the Fornax cluster. A detailed study to investigate or disentangle the contribution of NGC~1399 DM from the cluster potential is beyond the scope of this paper. However, our higher enclosed mass profile strongly suggests that by including ICGCs, we are also capturing the cluster DM potential. This finding indicates that ICGCs trace the cluster potential and play a crucial role in estimating the total mass of the system.

\begin{figure}[htp]
    \centering
    \includegraphics[width=1.05\linewidth]{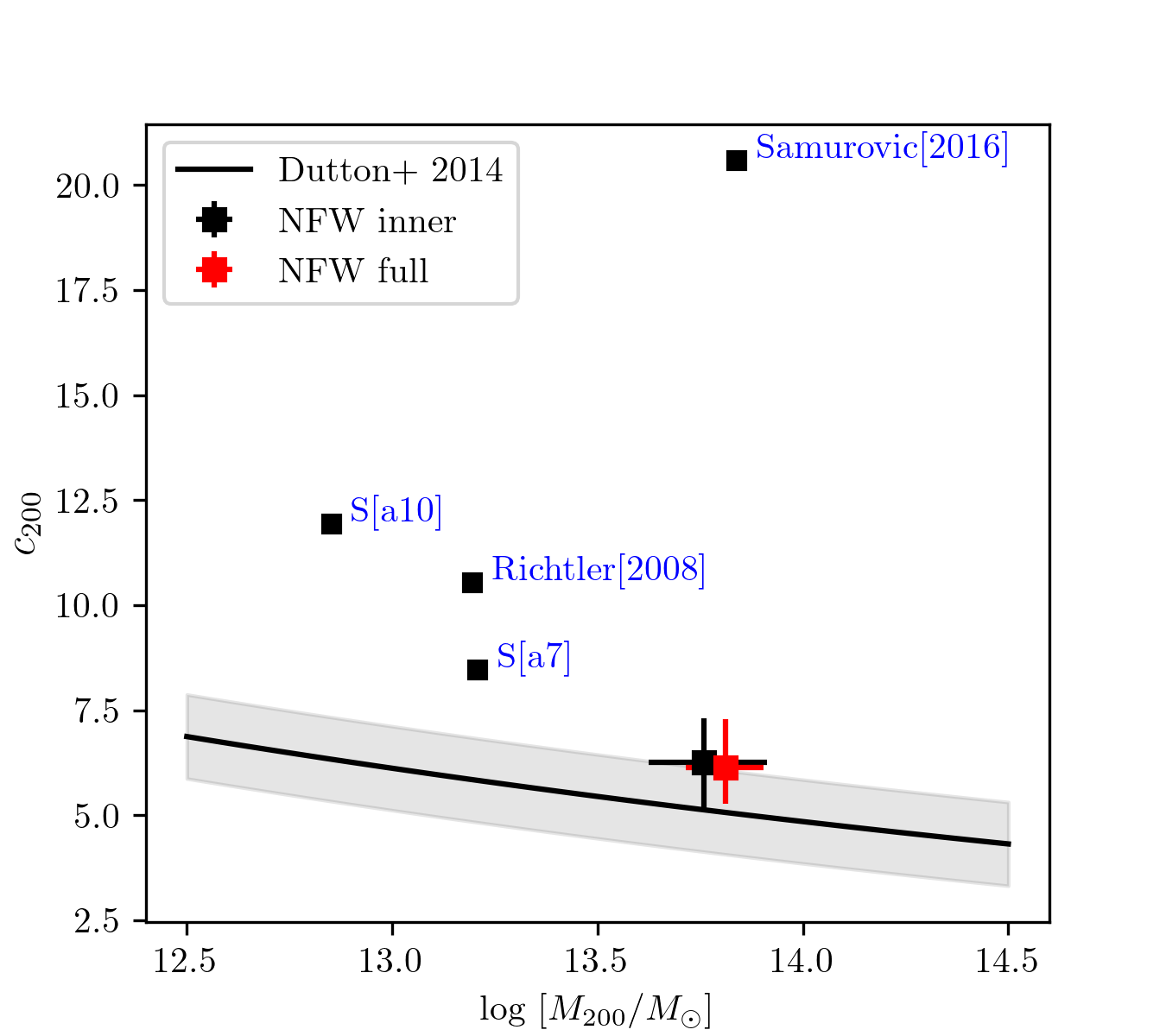}
    \caption{Virial mass and concentration parameter relation. The black line and shaded grey region indicate the $c_{200}$ and $M_{200}$ relation and its uncertainty from \cite{Dutton2014}. Open and filled red squares denote the measurements from our work for the \textit{inner} and \textit{full} sample. Black squares indicate the previously measured concentration parameter values of NGC~1399. The labels 'S[a7]', 'S[a10]', 'Richtler' and 'Samurovic' indicate the works from which these values are taken, from \cite{Schuberth2010}, \cite{ Richtler2008}, and \cite{Samurovic2016}, respectively.}
    \label{fig13:cvir_mvir}
\end{figure}

\subsection{Impact of the assumption of spherical symmetry}\label{sec5.2}
As discussed in Section~\ref{sec2:tracerdensity}, photometric studies of NGC~1399 show that it is not strictly spherical system. Its outer regions exhibit variations in ellipticity and position angle, which may indicate an intrinsically triaxial structure. In addition, the GC system itself shows mild projected flattening. In the present work, we assume spherical symmetry and negligible rotation in the GC system of NGC~1399. The latter assumption is supported by our rotation analysis in Section~\ref{sec2}, where we find that the GC system shows negligible rotation. However, it is important to assess whether the assumed spherical geometry could influence the inferred total mass profile. A full investigation of axisymmetric dynamical modelling, such as axisymmetric Jeans modelling \citep[e.g.][]{Dumont2024, versic2024}, would require a substantially more complex modelling framework than adopted here. Similarly, a fully triaxial dynamical analysis would require additional modelling machinery, likely including stellar kinematics and triaxial Schwarzschild modelling. Such an analysis is beyond the scope of the present work. Nevertheless, to test the impact of the assumed spherical symmetry on our inferred mass profile, we performed an additional consistency check in which all GC observables, including the tracer density and kinematic profiles, were re-extracted using elliptical annuli instead of the circular annuli adopted in the main analysis.

Figure~\ref{fig:c1} shows an example of the tracer density profile for the \textit{full} blue GC sample obtained using circular and elliptical annuli. Figure~\ref{fig:c2} compares the corresponding kinematic profiles for the red and blue GC populations in the \textit{full} sample. We then repeated the spherical Jeans modelling using the observables extracted from the elliptical annuli, for both the \textit{inner} and \textit{full} samples and for both halo parameterisations. We emphasise that, although the GC density and kinematic profiles were extracted using elliptical annuli, the corresponding dynamical models were still calculated using the spherical Jeans equations and did not adopt an ellipsoidal radius to account explicitly for the intrinsic flattening of the GC distribution. Thus, this test examines the effect of using different projected annular geometries rather than intrinsically flattened Jeans models. Despite the differences in the tracer density profiles, the inferred total mass profiles obtained from the circular and elliptical annuli are nearly indistinguishable within the $1\sigma$ uncertainties. Figure \ref{fig:c3} show the best-fitting models for the circular- and elliptical-annulus cases, while Figure \ref{fig:c4} shows the corresponding inferred mass profiles. 

Although the Jeans modelling itself remains spherical, this test modifies the input tracer profiles to explicitly account for the observed projected flattening of the GC system. Therefore, it directly probes the sensitivity of the inferred mass profile to the adopted projected geometry. The fact that the recovered mass profiles remain consistent within the uncertainties demonstrates that the spherical Jeans framework provides an adequate and robust description for the purposes of this study, despite the known departures from spherical symmetry in the tracer distribution. This further indicates that any bias introduced by the spherical assumption is sub-dominant compared to the statistical uncertainties of the data. This conclusion is also consistent with simulation-based studies of GC systems. For example, \cite{Hughes2021} investigated GCs as discrete kinematic tracers of the total mass distribution in Milky Way-mass galaxies and found that the robustness of the recovered mass profile is strongly affected by the number and radial distribution of tracers. Their results support the idea that, within the statistical limitations of discrete GC samples, the recovered total mass profile can remain robust even when simplified assumptions are adopted in the dynamical modelling.

\subsection{Comparison with simulation studies}

Here we compare our results with results from cosmological simulations in the $\Lambda$CDM framework. Several such simulations show a universal relation between DM concentration parameter $c_{vir}$ and virial mass \citep{Navarro1997, Bullock2001, Dutton2014, Schaller2015, Diemer2015}. The concentration parameter, $c_{vir}$, relates the inner scale radius with the virial radius and is expressed as $r_{vir}$/$r_{s}$. As mentioned in Section \ref{sec4}, for comparing the different models, we used $\Delta_{vir}$ $\sim$ 200. We use the concentration-mass relation from the work of \cite{Dutton2014}. They used the Planck cosmology \citep{Planck2014} in their DM-only cosmological simulations and derived the following halo-mass relation:

\begin{equation}
\rm log_{10}c_{200} = 0.905 - 0.101\log_{10}(M/10^{12} [M_{\odot}])
\end{equation}

In Figure \ref{fig13:cvir_mvir}, the black dashed line shows the relation between $c_{200}$ and $M_{200}$ of \cite{Dutton2014}, and the shaded grey region marks its 1$\sigma$ uncertainty. The filled and open red squares mark $c_{200}$ obtained for the NFW profile for the $\textit{full}$ and $\textit{inner}$ samples, respectively. The previous measurements of the concentration parameter of NGC~1399 are indicated as the black squares from the work of \cite{Schuberth2010, Richtler2008, Samurovic2016}. To enable a consistent comparison, we have rescaled these measurement to $\Delta_{200}$, however, obtaining their uncertainties was challenging. 

Our measured values of $c_{200}$ and $M_{200}$ for both \textit{inner} and \textit{full} sample are in good agreement with the simulation work based on the $\Lambda$CDM framework. The average concentration parameter recovered for the Fornax cluster is $\sim$ 6.20 indicating a low central density halo, consistent with expectations from cosmological simulations. One key reason for this agreement is the larger scale radius ($r_{s}$) obtained from our NFW fits for both samples, compared to previous studies. Our GC sample extends out to 150 kpc ($\sim$ 5 $\rm Re$) and the improved dynamical modelling method allowed us to trace the cluster-scale DM halo and robustly measure the scale radius of DM halo. This consistency between our measurements and cosmological predictions provides additional confidence to the robustness of our dynamically derived mass profiles.
 
\begin{figure}[htp]
    \centering
    \includegraphics[width=1.0\linewidth]{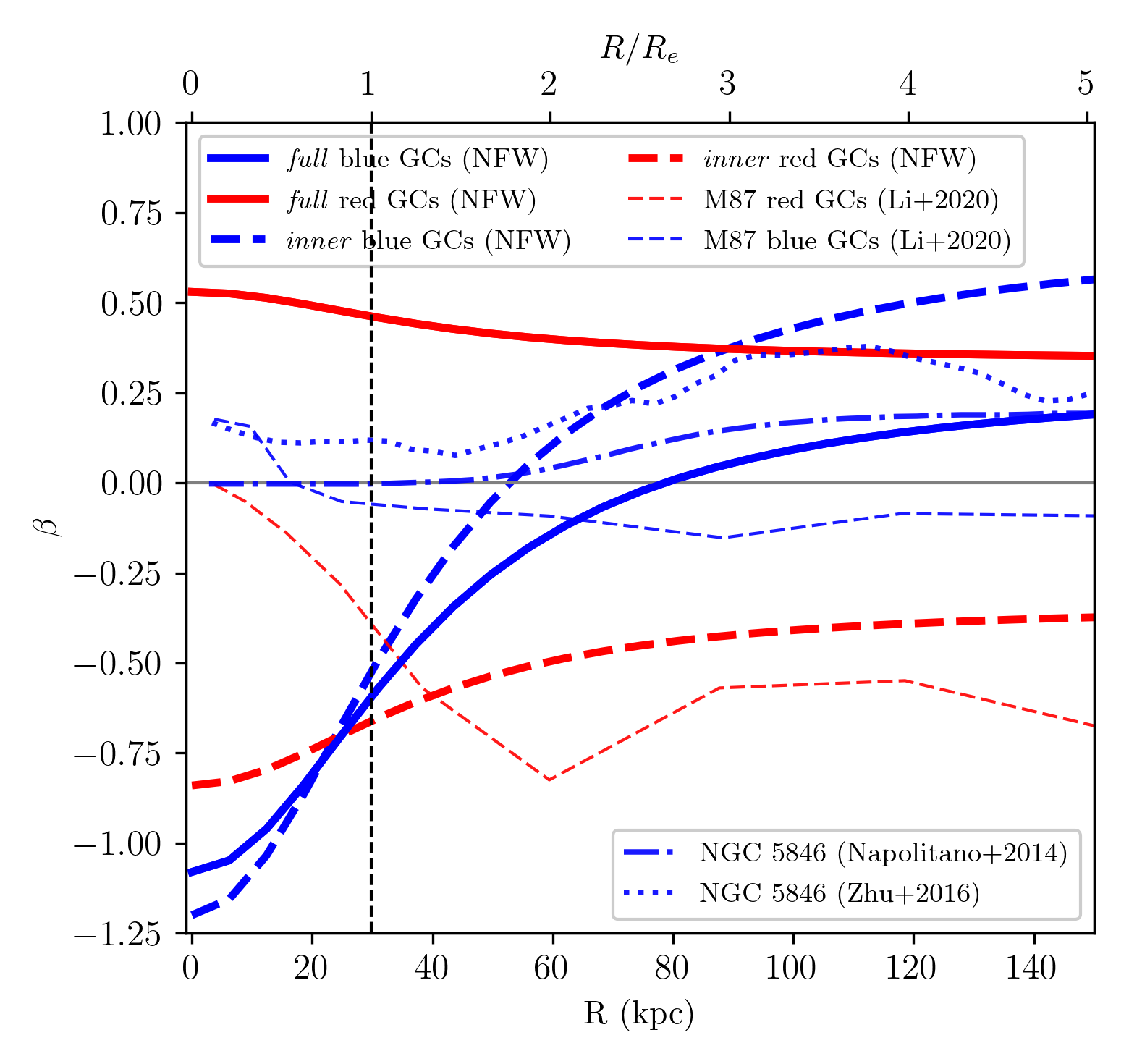}
    \caption{Velocity anisotropy distribution of GCs in the \textit{full} sample. Solid red and blue lines indicate the anisotropy for the red and blue GCs for the NFW halo. Dashed red and blue lines indicate the velocity anisotropy of M87 from \cite{Chao2020}. Light blue and dark blue dashed lines indicate the anisotropy profiles of NGC\,5846's blue GCs adopted from \cite{Zhu2016} and \cite{napolitano2014}, respectively. }
    \label{fig:anisotropy}
\end{figure}

\subsection{GCs orbital distribution}

Finally, we compare our GCs anisotropy profiles with those of other massive systems available in the literature and discuss the implication of our results. Figure \ref{fig:anisotropy} shows the orbital velocity anisotropy profile obtained for the \textit{full} sample using the NFW halo. Since the orbital anisotropy is similar for both the Burkert and NFW halos, we plot only the profile obtained from the NFW halo. The solid blue and red curves denote the anisotropy profiles of the \textit{full} blue and red GC samples, respectively, while the thick dashed blue and red curves represent the corresponding profiles for the \textit{inner} samples.

We find that red GCs show a relatively smooth anisotropy profile with radius. For the \textit{full} sample, the red GCs exhibit mildly radial anisotropy over the full radial range, with $\beta \sim 0.3 - 0.5$, consistent with a nearly isotropic orbital distribution. In contrast, the \textit{inner} red GC sample exhibits tangential anisotropy throughout the radial range, although the anisotropy gradually increases towards larger radii. The blue GCs exhibit a more complex orbital structure compared to the red GCs. For both the \textit{full} and \textit{inner} blue GC samples, we find strong tangential anisotropy in the inner regions ($R \lesssim 2R_{\rm e}$), with $\beta \lesssim -1$ near the centre. However, the anisotropy profile increases strongly with radius and transitions to radial anisotropy at larger radii. Beyond $\sim2R_{\rm e}$, the \textit{full} blue GC sample becomes mildly radial, while the \textit{inner} blue GC sample shows strongly radial anisotropy with $\beta \gtrsim 0.5$ in the outskirts. The transition from tangential to radial anisotropy suggests that the outer blue GCs preferentially occupy radial orbits, indicating that a significant fraction of these objects may have been accreted through radial assembly processes within the Fornax cluster environment.

Previously, \cite{Saglia2000} studied the stellar kinematics of NGC~1399 using integrated-light spectroscopy and found positive anisotropy values within $\sim1R_{\rm e}$. However, the comparison between the stellar anisotropy profile and our GC-based anisotropy profile is not strictly one-to-one because the stellar field population and the GC subpopulations probe dynamically distinct tracers with different spatial distributions and assembly histories. In addition, previous Jeans modelling studies of the NGC~1399 GC system by \cite{Schuberth2010} highlighted the strong mass--anisotropy degeneracy and the difficulty of obtaining a unique orbital solution for both the red and blue GCs simultaneously. In particular, \cite{Schuberth2010} showed that a wide range of anisotropy assumptions, ranging from mildly tangential to radial orbits, could reproduce the observed GC kinematics, especially for the blue GCs which exhibit more complex dynamical behaviour.

In the literature, few studies have explored the orbital distribution of GCs around massive galaxies \citep{Agnello2014, napolitano2014, Zhu2016, Chao2020}. These studies are based on different sample sizes and methodologies, leading to variations in the measured anisotropy, even for the same system. Considering these caveats, we compare our results with those for massive systems such as NGC\,5846 and M87 and draw conclusions based on recent simulation studies by \cite{Ramos2018}.

M87 is the central massive galaxy of the Virgo Cluster and hosts a rich GC system. \cite{Agnello2014} found that red GCs around M87 exhibit slightly tangential orbits in the inner regions and a nearly isotropic behaviour in the outskirts. In contrast, the blue GCs are nearly isotropic in the centre but show mild tangential anisotropy at larger radii (>100 kpc). More recently, \cite{Chao2020} studied the M87 GC system out to 400 kpc using discrete dynamical modelling. They found that red GCs are isotropic in the centre but become tangential in the outskirts, while blue GCs are radially anisotropic in the inner regions and slightly tangential in the outskirts. In Figure \ref{fig:anisotropy}, the dashed blue and red lines denote the velocity anisotropy profiles of M87 from \cite{Chao2020}. In contrast, we find that in the outskirts of NGC~1399, both red and blue GCs exhibit mild and strong radial anisotropy, respectively. The light blue and dark blue dotted curves in Figure \ref{fig:anisotropy} show the anisotropy profiles of NGC\,5846 from \cite{Zhu2016} and \cite{napolitano2014}, which are consistent with our findings for blue GCs in the outer regions of NGC~1399.

Our finding of radial anisotropy of the outer GCs, particularly the blue GCs, is consistent with the results from the Illustris simulation. \cite{Ramos2018} found that ICGCs exhibit radial anisotropy ($\beta$ $\geq$ 0.3), suggesting that these GCs have been accreted on radial orbits. Our modelling results support an accretion scenario for blue GCs, which aligns with previous photometric and spectroscopic studies of blue GCs \citep{Peng2006}. In the future, it will be interesting to investigate how the inclusion of inner stellar kinematics of NGC~1399 affects the dynamical modelling and orbital distributions of the inner red and blue GCs.

%--------------------------------------------------------------------
\section{Summary and Conclusions}\label{sec6}

In this work, we have applied spherical, non-rotating Jeans modelling to perform the dynamical mass-modelling of NGC~1399 using its globular cluster (GCs) system out to 150 kpc. For this, we used the FVSS GCs radial velocity catalogue produced in the FVSS-III paper. To investigate the role of intra-cluster GCs in the mass-modelling, we divided the full radial velocity catalogue into two samples: The \textit{inner} sample, consisting of GCs within 2.5$\rm Re$ ($\sim$ 60 kpc) of NGC~1399 and the \textit{full} sample, which extends to 5$\rm Re$ ($\sim$ 150 kpc) and includes intra-cluster GCs. 

For mass-modelling work, we solved the Jeans equation to fit the projected velocity dispersion and kurtosis of the observed GCs, obtaining the total mass of NGC~1399 and velocity anisotropy of the GCs system. Our approach involves jointly modelling both red and blue GCs which helps in alleviating the mass-anisotropy degeneracy. For the dark matter (DM) halo, we considered both a cuspy (NFW) profile and cored (Burkert) halo. 

The following are the main results of our work:

1) We find that, for both the \textit{inner} and \textit{full} samples, both NFW (cuspy) and Burkert (cored) DM halos reproduce the observed kinematics well (Figures \ref{fig:best_fit_param_NGC~1399} and \ref{fig:best_fit_param_NGC~1399+IC}). Since both halo profiles provide similarly good fits to the data, we cannot conclusively determine whether a cuspy or cored profile is preferred.

2) We assumed a constant stellar mass-to-light ratio, $\Upsilon_{*,g}$, in our mass modelling. The best-fitting $g$-band values are $\Upsilon_{*,g} = 4.04^{+1.30}_{-1.25}$ and $3.35^{+1.11}_{-1.02}$ for the \textit{inner} sample, and $\Upsilon_{*,g} = 6.08^{+1.76}_{-1.88}$ and $4.59^{+1.85}_{-1.87}$ for the \textit{full} sample, for the Burkert and NFW haloes, respectively. After accounting for differences in photometric bands, these values are broadly consistent with previous estimates reported in the literature \citep{Gebhardt2007,Vaughan2018} (Section~\ref{sec4.3}).

3) The total enclosed mass profiles obtained from the Burkert and NFW halos agree well within the observed radial range for both the \textit{inner} and \textit{full} samples. Our enclosed mass profile is also consistent with previous measurements of NGC~1399 within $1\rm Re$. Independent of the adopted DM halo profile, the mass budget is DM-dominated, with the DM fraction exceeding 50\% within the \textit{inner} sample at $1\rm Re$ and rising to more than 90\% at $5\rm Re$ ($\sim$145 kpc).

4) Based on the best-fit parameters of our models, we measured $M_{200}$ for NGC~1399 and find that the Burkert and NFW halos imply different virial masses for both samples. For the \textit{full} sample, the virial mass is $\log M_{200}=13.49^{+0.06}_{-0.05},M{_\odot}$ for the Burkert halo, whereas for the NFW halo it is $\log M_{200}=13.81^{+0.09}_{-0.09},M{\odot}$ (Table \ref{Tab:best_fit_params}). Thus, the NFW halo favours a systematically larger virial mass than the Burkert halo, although the difference between the two models is less extreme for the \textit{full} sample than for the \textit{inner} sample. For the NFW halo, these values lie close to the expected $c_{200}$--$M_{200}$ relation from cosmological simulations (Figure \ref{fig13:cvir_mvir}).

5) We studied the GC velocity anisotropy profiles using the anisotropy parametrisation adopted in the dynamical modelling. For the \textit{inner} sample, we found that the red GCs are tangentially anisotropic within $\sim$10 arcmin, whereas the blue GCs show tangential anisotropy within the inner $\sim$15 arcmin and gradually become radially anisotropic at larger radii. For the \textit{full} sample, the red GCs remain radially anisotropic over the full radial range, while the blue GCs show strong tangential anisotropy within $\sim$40 kpc, become mildly tangential out to $\sim$80 kpc, and are approximately isotropic or mildly radially anisotropic in the outskirts (Figure \ref{fig:anisotropy}). The radial anisotropy of the blue GCs in the outskirts supports an accreted or infalling origin for these GCs within the Fornax cluster potential.

Our results demonstrate that including ICGCs, which extend further into the outskirts of the system, allows us to trace the cluster potential more effectively. Incorporating ICGCs in our mass modelling therefore enables a robust measurement of the total enclosed mass profile of NGC~1399. In future work, we aim to quantify the impact of the substructures observed around NGC~1399 and move beyond the spherical modelling using upcoming FVSS survey data. We also plan to perform a multi-component mass modelling of the Fornax cluster, incorporating the stellar kinematics of NGC~1399, to gain deeper insights into the mass assembly history of the cluster.

\small

\vspace{3.0mm}

%Thanks to the anonymous referee for helpful feedback and suggestions that improved the manuscript's scientific content. We would also like to thank the language editor for improving the text of the article

\textbf{Acknowledgements:} We thank the anonymous referee for the helpful and insightful scientific comments, which improved the content and clarity of the manuscript. A. Chaturvedi thanks Tom Richtler, Eric Emsellem, Lodovico Coccato, Jens Thomas and Davor Krajnovi\'c for helpful discussions. M.H. acknowledges financial support from the Excellence Cluster ORIGINS which is funded by the Deutsche Forschungsgemeinschaft (DFG, German Research Foundation) under Germany’s Excellence Strategy – EXC 2094 – 390783311. K.F. acknowledges support from the European Union’s Horizon 2020 research and innovation programme under the Marie Sk\l{}odowska-Curie grant agreement No 101103830. M.C. acknowledges support from the ASI-INAF agreement “Scientific Activity for the Euclid Mission” (n.2024-10-HH.0; WP8420) and from the INAF “Astrofisica Fondamentale” GO-grant 2024. This research made use of various software including: Astropy (\href{https://www.astropy.org/}{https://www.astropy.org})-a community-developed core Python package for Astronomy \citep{Astropy2013, Astropy2018}, pPXF \citep{Cappellari2017}, scipy \citep{Scipy2020}, emcee \citep{emcee2013}.

\bibliographystyle{aa} % style aa.bst
\bibliography{aa55529-25}

\begin{appendix}

\section{Jeans formalism and mass modelling details}
\subsection{Jeans modelling}\label{Jeans_model}

The Jeans equations are obtained by taking the velocity moments of the collisionless Boltzmann equation, which describes the phase-space distribution of tracers in a given gravitational potential. The Jeans equation connects the tracer velocity distribution and its anisotropy to the total dynamical mass of the system. Assuming a spherical symmetry and a non-rotating system, the Jeans equation is written as:

\begin{equation} \label{equ5}
    \frac{d}{dr}(j\sigma^{2}_{r}) + \frac{2\beta}{r}j\sigma^{2}_{r} = -j\frac{d\phi}{dr} = j\frac{G  M_{\rm tot}}{r^2}
\end{equation}

Here $j$ is the tracer 3D number density, $M_{\rm tot}$ is the total mass of the system, $\sigma_{r}$ is the radial velocity dispersion, and $\phi$ is the gravitational potential and $\beta$ is the tracer velocity anisotropy, expressed as:

\begin{equation} \label{equ6}
\beta = 1 - \frac{\sigma_{t}^{2}}{\sigma_{r}^{2}}
\end{equation}

where $\sigma_t$ is the tangential velocity dispersion, for which we assume
%Here, 
$\sigma_{t}$=$\sigma_{\phi}$=$\sigma_{\theta}$, being $\sigma_{\phi}$ and $\sigma_{\theta}$, the components of the velocity dispersion tensor along the two angular coordinates of a spherical coordinate system.

One of the limitations of the Jeans modelling technique is the mass-anisotropy degeneracy. To reduce this degeneracy, the approach is to assume and consider different orbital distributions like radial, tangential or isotropic and solving equation \ref{equ5} to obtain the projected velocity dispersion. Additional constraints on the mass-anisotropy degeneracy can be obtained by including the fourth moment of the LOSVD, which contains information about the orbital distribution of the tracers.
For a discrete kinematical sample a good estimator of the fourth moment of the LOSVD is the kurtosis \citep[see e.g.][]{Lokas2003}.
The projected fourth-moment Jeans equations writes:

\begin{equation} \label{jeans_equ2}
    \frac{d}{dr}(j\overline{v_{r}^{4}}) + \frac{2\beta}{r}j\overline{v_{r}^{4}} = -3j\sigma^{2}_{r}\frac{d\phi}{dr}. 
\end{equation}

Using this equation to constrain the kurtosis can alleviate and control the mass-anisotropy degeneracy \citep[see][]{Lokas2003, Lokas2005}. 
This procedure has been applied and well tested for the dynamical modelling of massive galaxies \citep[for details see][]{napolitano2009,  napolitano2011, napolitano2014}. 

We apply the dispersion-kurtosis analysis and use the solution of the Jeans equations, $\sigma_r$ and $\overline{v_r^4}$,
to obtain the projected second and fourth velocity moments as:

\begin{equation} \label{equ7}
    \sigma^{2}_{los}(R) = \frac{2}{I(R)} \int_{R}^{\infty} \left(1 - \beta\frac{R^{2}}{r^{2}}\right) \frac{j \sigma^{2}_{r}}{\sqrt{r^{2}-R^{2}}}dr
\end{equation}

and projected fourth moment, 

\begin{equation} \label{equ8}
    v^{4}_{los}(R) = \frac{2}{I(R)} \int_{R}^{\infty} \left(1 - 2\beta\frac{R^{2}}{r^{2}} + \frac{\beta(1+\beta)}{2}\frac{R^{4}}{2r^{4}}\right) \frac{j\overline{v_{r}^{4}}r}{\sqrt{r^{2}-R^{2}}}dr
\end{equation}

In Equation \ref{equ8}, $\beta$ is constant, but can be generalised for radial dependence \citep[see Eq. 35 and 37 of][]{Richardson2013}. In our case, we consider the $\beta$ parametrisation introduced by \cite{churazov2010} :

\begin{equation} \label{equ9}
 \beta(r) = \frac{\beta_{in}r_{a}^c+\beta_{out}r^{c}}{r^{c} + r_{a}^c}    
\end{equation}

Equation \ref{equ9} is characterised by the two anisotropy parameters $\beta_{in}$ and $\beta_{out}$ that produce asymptotic values for $r\to 0$ and $r\to\infty$, respectively. The parameter $r_{a}$ represents the anisotropy radius, while the exponent '$c$' regulates the transition between $\beta_{in}$ and $\beta_{out}$. 
%We adopt $c=2$,  which corresponds to the Osipkov–Merritt model \citep{Osipkov1979, Merritt1985}, where  $\beta_{in}=0$ and $\beta_{out}=1$. For our modelling purpose, we fix $c=2$ and vary $\beta_{in}$ and $\beta_{out}$. 
The Osipkov–Merritt model \citep{Osipkov1979, Merritt1985} corresponds to $c=2$ with $\beta_{in}=0$ and $\beta_{out}=1$. For our modelling purpose, we adopt $c=2$, but vary $\beta_{in}$ and $\beta_{out}$. 
Regarding the anisotropy radius, we tested different values between 1–2 $\rm R_{e}$ of NGC~1399, as this is the region where we expect a transition in the anisotropy profile, as seen in the total velocity dispersion profile of GCs \citep{Chaturvedi2022} and also noticed in the photometric observations \citep{Iodice2019}. To reduce the number of free parameters in the model, we fix $r_{a}$ at 40 kpc.

\begin{figure}
    \centering
    \includegraphics[width=1.05\linewidth]{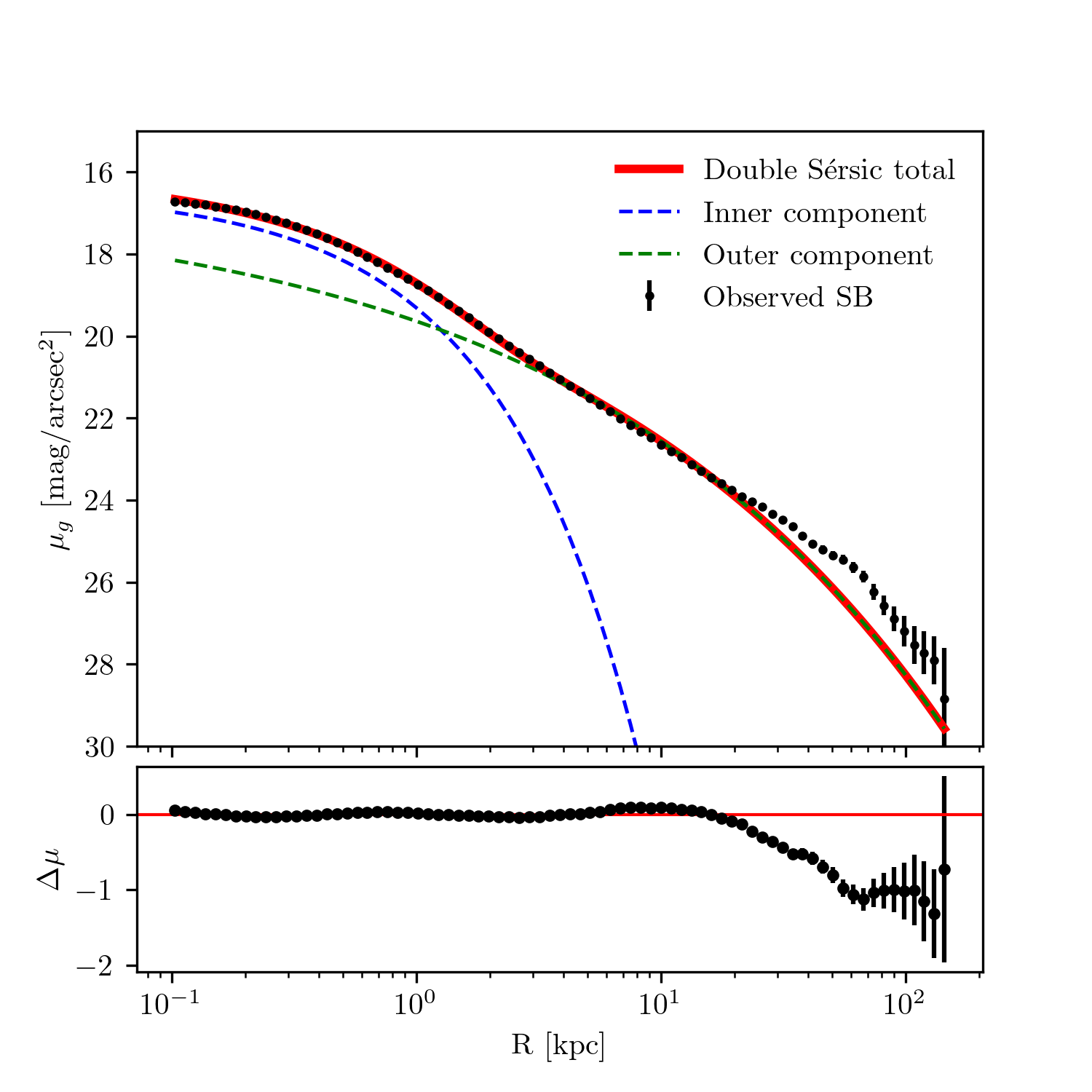}
    \caption{Double S\'ersic profile fit to the observed FDS g-band surface brightness profiles of NGC~1399. The g-band surface bright of NGC~1399 is adopted from \cite{Iodice2016}.}
    \label{fig:NGC~1399_sb_fit}
\end{figure}

\begin{table}[h]
\caption[]{Best fit parameters of the S\'{e}rsic profile for the NGC 1399 surface brightness profile}
\renewcommand{\arraystretch}{1.15}
\label{Taba1:NGC~1399_sb}
\centering
\begin{tabular}{ccccccc}
\hline

\textbf{Sample} & $\mu_{e}$ & $R_{e}$ & $n_{e}$  \\ 
\textit{inner}   & [$mag/arcsec^{2}$]  & [kpc]   \\ 
\hline
    % All   & 0.063 & 40.39 & 2.11  \\
    Inner component   & 18.99 & 0.85 & 1.32  \\
    Outer component   & 23.69 & 18.10 & 3.44  \\
\hline
\end{tabular}
\end{table}

\subsection{Gravitational potential}
The modelled gravitational potential in our dynamical modelling consists of two main components: stellar mass and DM halo mass. The total mass as in Eq. \ref{equ5}, is defined as:
\begin{equation}
   M_{tot}(r) = M_{*}(r) + M_{DM}(r)
\end{equation}

where $M_{*}$ and $M_{DM}$ are the stellar and DM mass.

The stellar mass profile, \(M_{*}(r)\), is obtained by integrating the deprojected \(g\)-band luminosity profile, \(\nu_{*}\), of NGC~1399, adopted from \citet{Iodice2016}. Because the FDS \(g\)-band surface-brightness profile of NGC~1399 extends beyond 150~kpc, we modelled it using a double-S\'ersic fit, which allows for easy de-projection of projected S\'ersic profile using Abel integration. Figure \ref{fig:NGC~1399_sb_fit}, shows double-S\'ersic fit to the g-band surface brightness profile of NGC~1399 and Table \ref{Taba1:NGC~1399_sb} shows the two component S\'{e}rsic profile parameters. The resulting stellar mass profile, \(M_{*}(r)\), is expressed as:

\begin{equation}
    M_{*}(r) = 4\pi\Upsilon_{*} \int_{0}^{r} \nu_{*}(r)r^{2}dr
\end{equation}

where $\Upsilon_{*}$ is the stellar mass-to-light ratio, which is a free parameter in the modelling.

\subsubsection{Dark matter halo}

Numerical DM only simulations have shown that the DM halo can be well described with a cuspy profile in the centre. However, the very inner shape of the profile depends on the numerical details used in the simulations \citep{Bullock2001, Schaller2015}. Observational studies of low-surface brightness galaxies, have shown a cored DM profile \citep{Burkert1995}. To properly investigate what type of DM halo profile is preferred for the modelled system, whether it's cuspy or cored, it is important to include the stellar kinematics of the galaxy. In a future work, we plan to incorporate the stellar kinematics of NGC~1399 along with the GCs kinematics into our dynamical modelling and consider the generalized Navarro–Frenk–White profile, which provides a more comprehensive approach to infer the inner slope of DM halo profile. 

For this work,  we use a core and cuspy DM halo profile parameterized as  Burkert profile \citep{Burkert1995} and NFW profile \citep{Navarro1997}, respectively. The Burkert DM density profile is characterised by the scale radius $r_{s}$ and density normalisation parameter $\rho_{o}$, and is expressed as follows:
\begin{equation}
    \rho(r) = \frac{\rho_{o}}{\left(1 + \frac{r}{r_{s}} \right)\left(1 + \frac{r^{2}}{r_{s}^{2}} \right)}
\end{equation}

The cumulative mass profile for the Burkert halo is given as:

\begin{equation}
    M(r) = 4\pi\rho_{o} r_{s}^{3}\left( \frac{1}{2}ln\left(1 +\frac{r}{r_{s}}\right)  + \frac{1}{4}ln\left(1 +\frac{r^{2}}{r_{s}^{2}}\right) - \frac{1}{2}\text{arctan}\left(\frac{r}{r_{s}} \right)\right)
\end{equation}

% \subsubsection{NFW profile}
The NFW profile is given by:

\begin{equation}
    \rho_{NFW}(r) = \frac{\rho_{o}}{\left(\frac{r}{r_{s}} \right)\left(1 + \frac{r}{r_{s}} \right)^{2}}
\end{equation}

where $\rho_{s}$ and $r_{s}$ are the characteristic density and radius. The cumulative mass profile for the NFW halo is expressed as:

\begin{equation}
    M_{NFW}(r) = 4\pi\rho_{o} r_{s}^{3}\left(ln\left(1 +\frac{r}{r_{s}}\right)  - \frac{\frac{r}{r_{s}}}{1 +\frac{r}{r_{s}}}\right)
\end{equation}

\section{Corner plots showing posterior distribution of parameters} 
\begin{figure*}
    \centering
    \includegraphics[width=19cm]{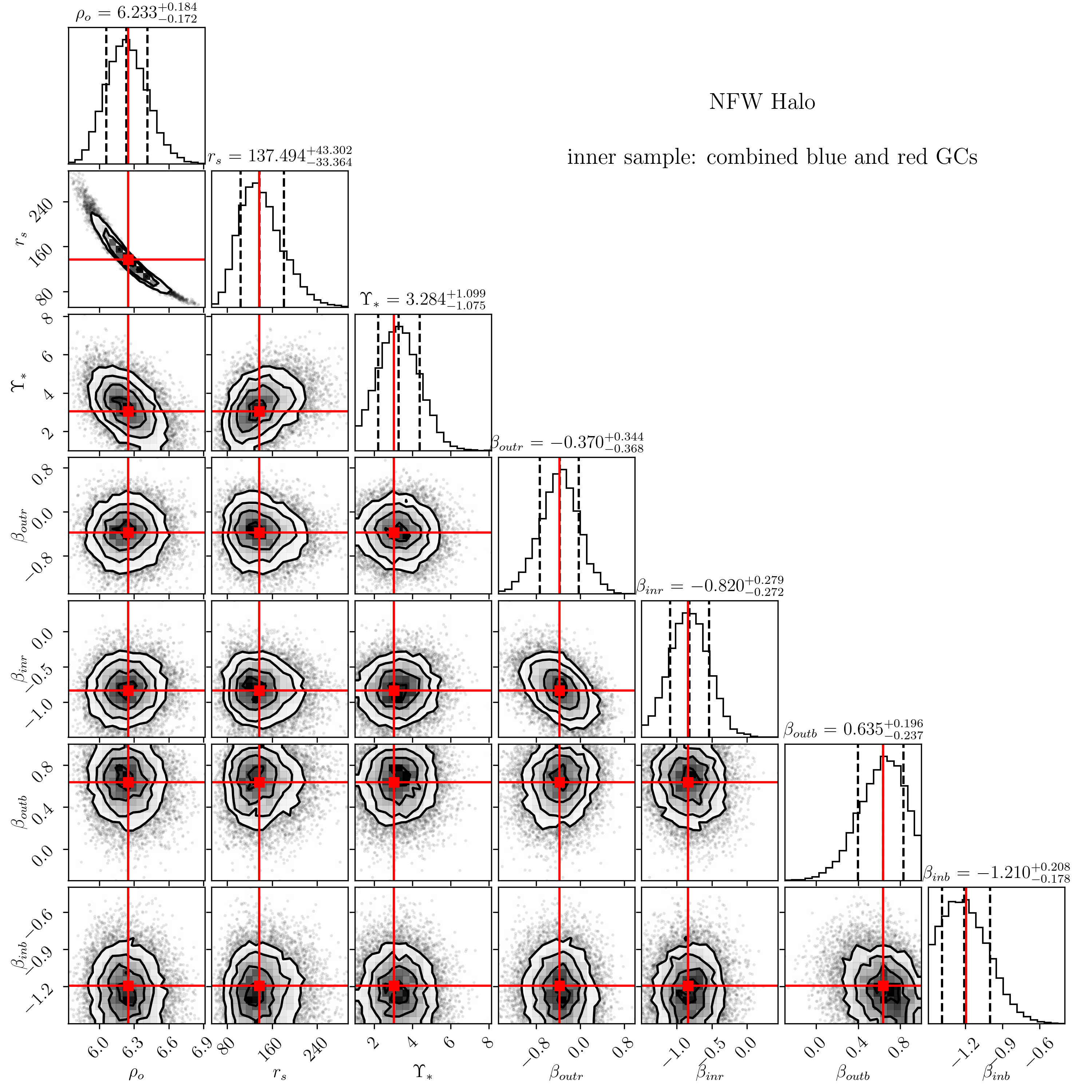}
    \caption{1D and 2D posterior distribution of parameters from the two-component Jeans modelling using the NFW halo for the
\textit{inner} sample. Rest same as Figure \ref{fig:bur_NGC~1399_corner}.}
    \label{fig:nfw_NGC~1399_corner}
\end{figure*}

\begin{figure*}
    \centering
    \includegraphics[width=19cm]{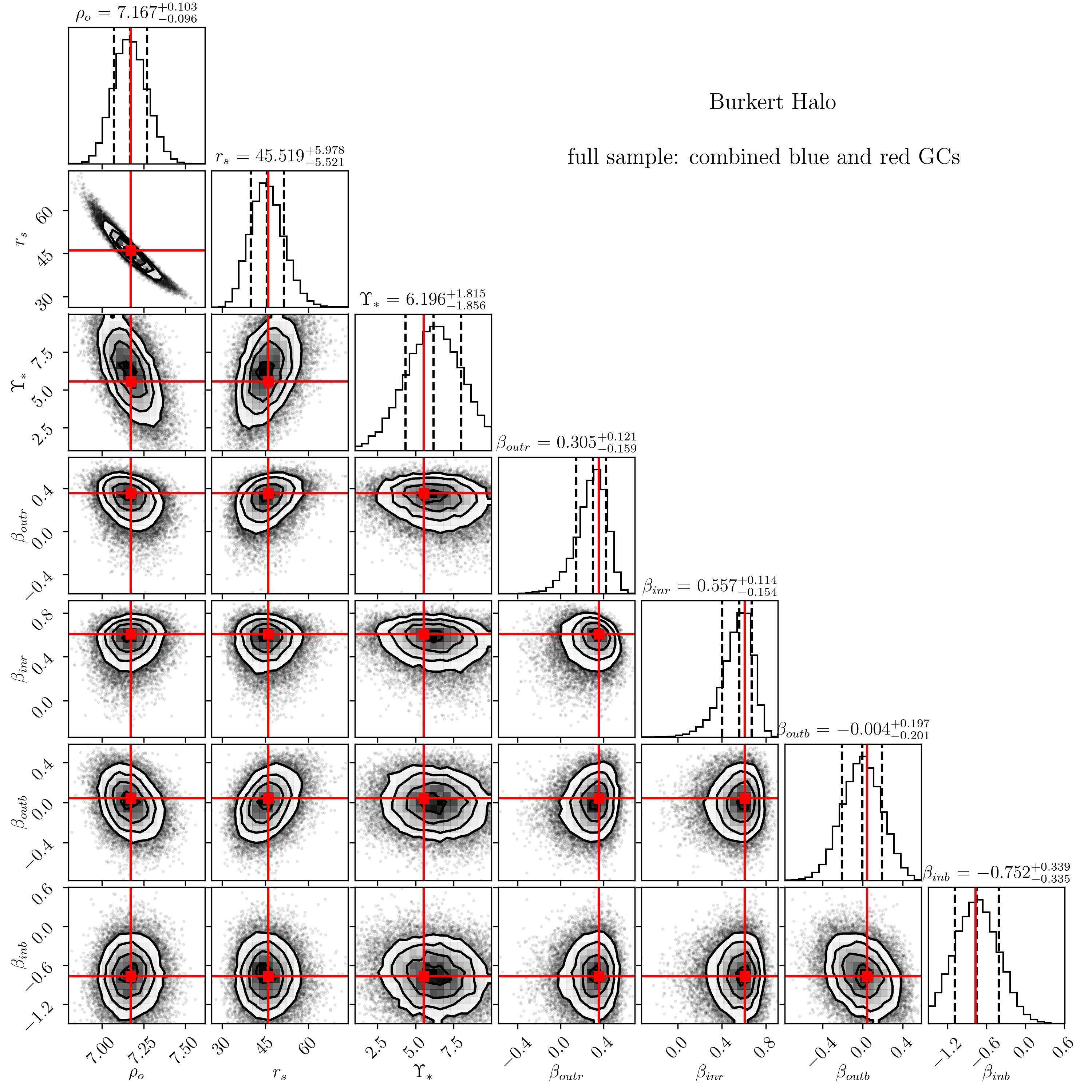}
    \caption{1D and 2D posterior distribution of parameters from the two-component Jeans modelling using the Burkert halo for the
\textit{full} sample. Rest same as Figure \ref{fig:bur_NGC~1399_corner}.}
    \label{fig:bur_NGC~1399+IC_corner}
\end{figure*}

\begin{figure*}
    \centering
    \includegraphics[width=19cm]{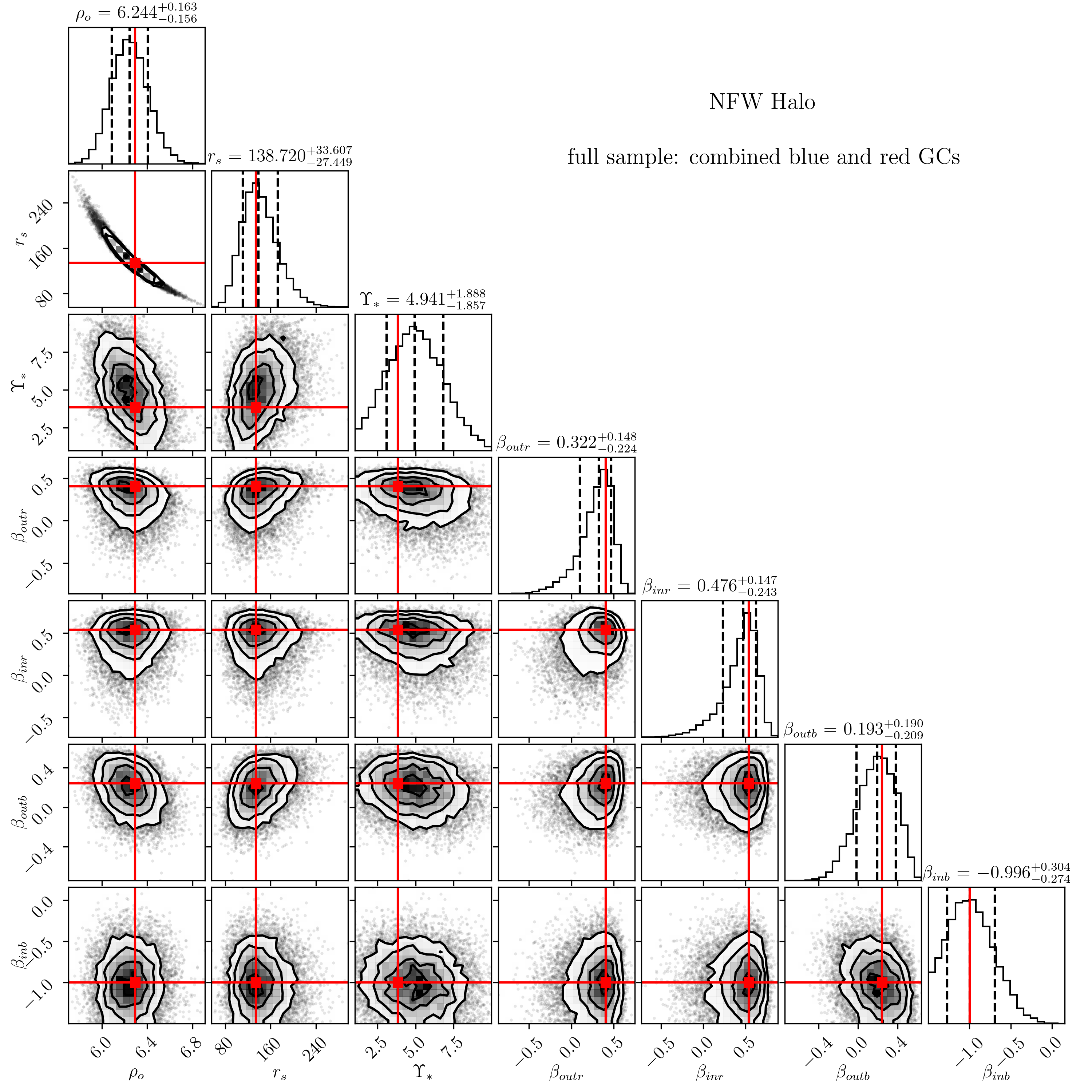}
    \caption{1D and 2D posterior distribution of parameters from the two-component Jeans modelling using the NFW halo for the
\textit{full} sample. Rest same as Figure \ref{fig:bur_NGC~1399_corner}.}
    \label{fig:nfw_NGC~1399+IC_corner}
\end{figure*}

In this appendix we show the 1D and 2D posterior distribution of parameters (MCMC output) from the two-component Jeans modelling for the \textit{inner} and \textit{full} sample. Figure \ref{fig:nfw_NGC~1399_corner} shows the 1D and 2D corner plots for the NFW halo for the inner sample where as Figure \ref{fig:bur_NGC~1399+IC_corner} and Figure \ref{fig:nfw_NGC~1399+IC_corner} shows the same but for Burkert and NFW halo for the \textit{full} sample.

% FIGURE 7

%### FIGURE 10 ###

\section{Circular versus elliptical annuli for the GC system}
In this appendix, we present the figures associated with the test described in Section~\ref{sec5.2}. We compare the GC observables extracted using circular and elliptical annuli, focusing on the \textit{full} sample. The kinematic comparison is shown for the red and blue GC populations, while the corresponding Jeans-model fits are shown for the NFW halo case. We have also performed the same test for the Burkert halo and find consistent results.

Figure~\ref{fig:c4} shows the inferred total mass profiles for the \textit{full} sample obtained using circular and elliptical annuli, for both the Burkert and NFW halo models. The mass profiles agree within the uncertainties, demonstrating that the inferred mass distribution is insensitive to the adopted annulus geometry.

\begin{figure*}
    \centering
    \includegraphics[width=0.90\linewidth]{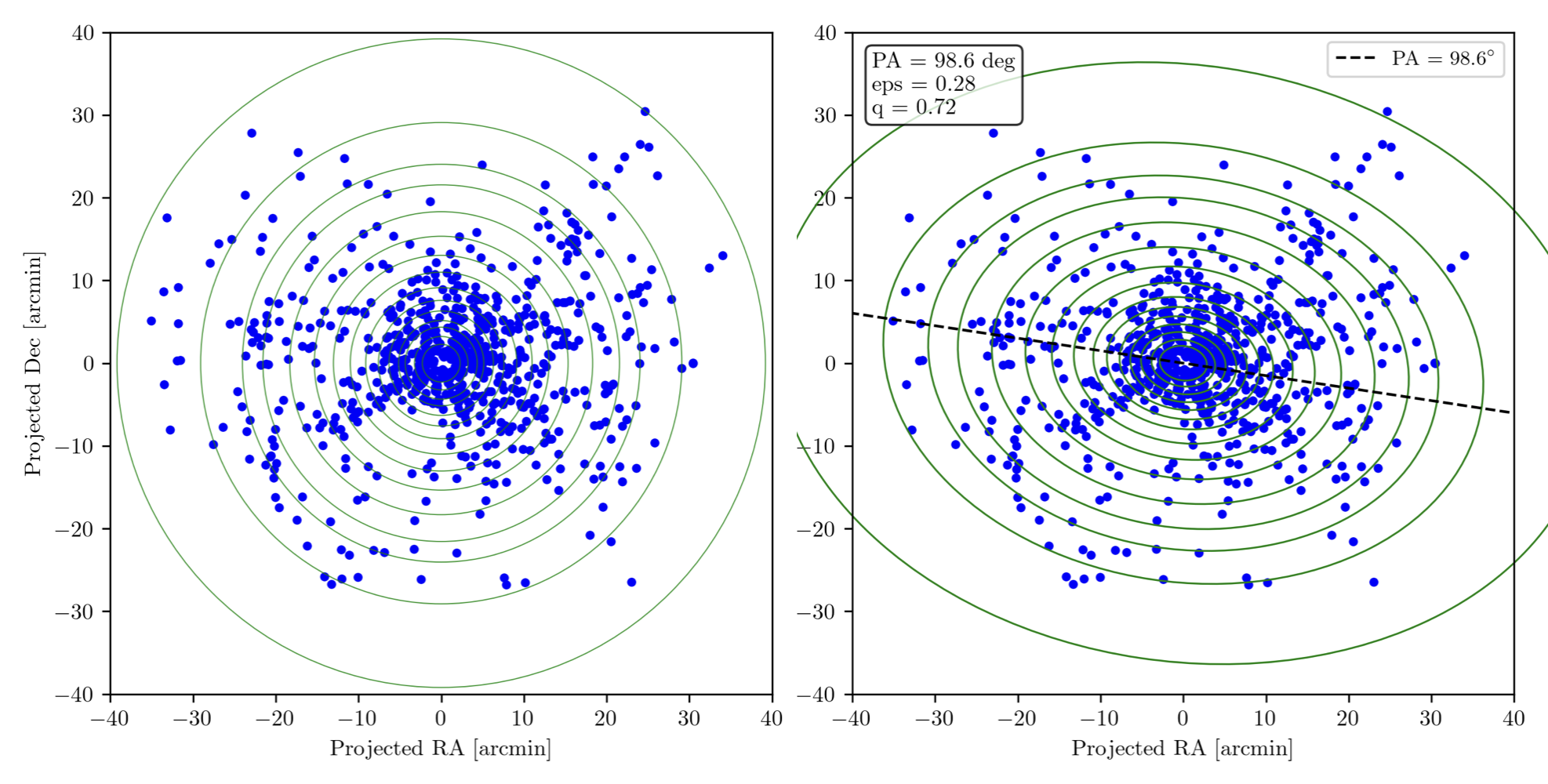}
    \caption{Example of circular (left) and elliptical (right) annuli used to extract the tracer density of blue GCs for the \textit{full} sample. The black dashed line in right panel shows the position angle (PA) of the elliptical annuli. }
    \label{fig:c1}
\end{figure*}

\begin{figure*}
    \centering
    \includegraphics[width=0.90\linewidth]{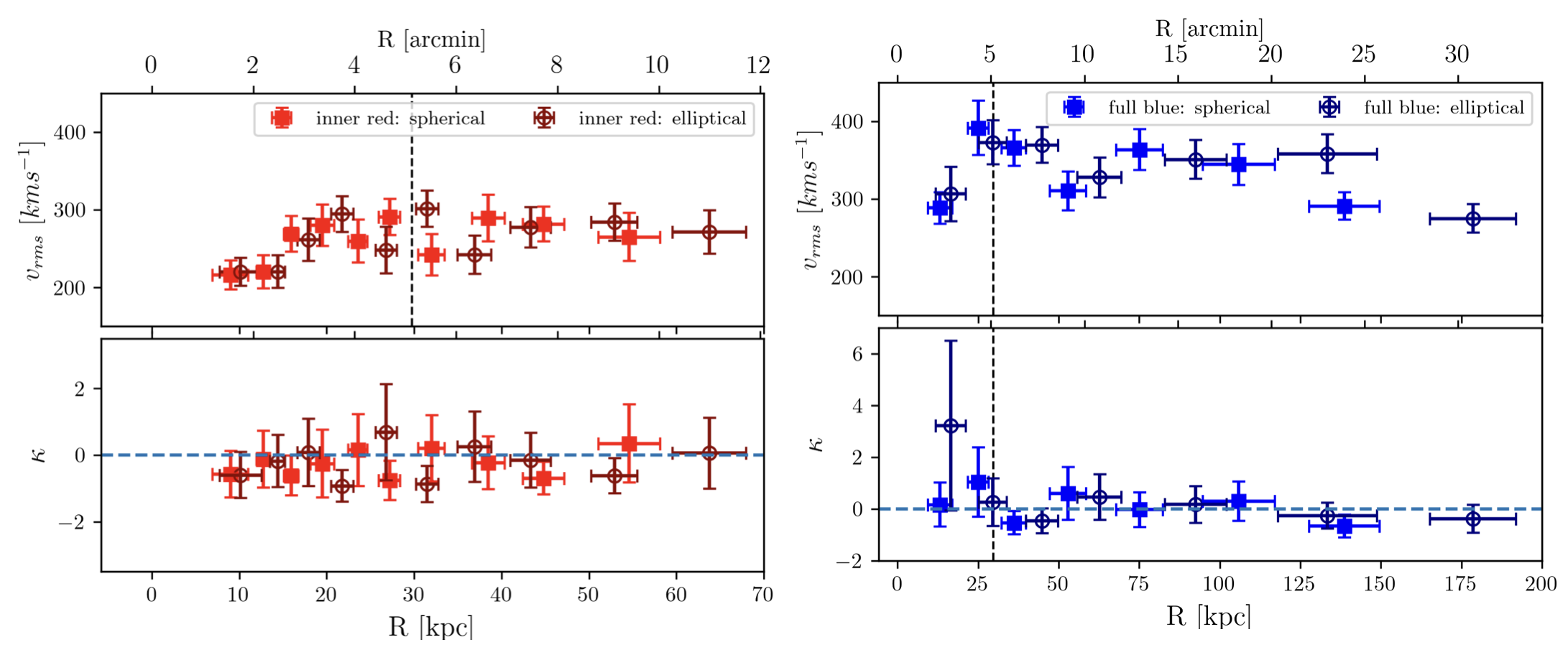}
    \caption{Comparison of the kinematic profiles of the red (left) and blue (right) GCs in the \textit{full} sample, extracted using circular and elliptical annuli. The velocity-dispersion and kurtosis profiles obtained with the two approaches agree within the uncertainties, indicating that the extracted GC kinematics are insensitive to the adopted annulus geometry.}
    \label{fig:c2}
\end{figure*}

\begin{figure*}
    \centering
    \includegraphics[width=18cm]{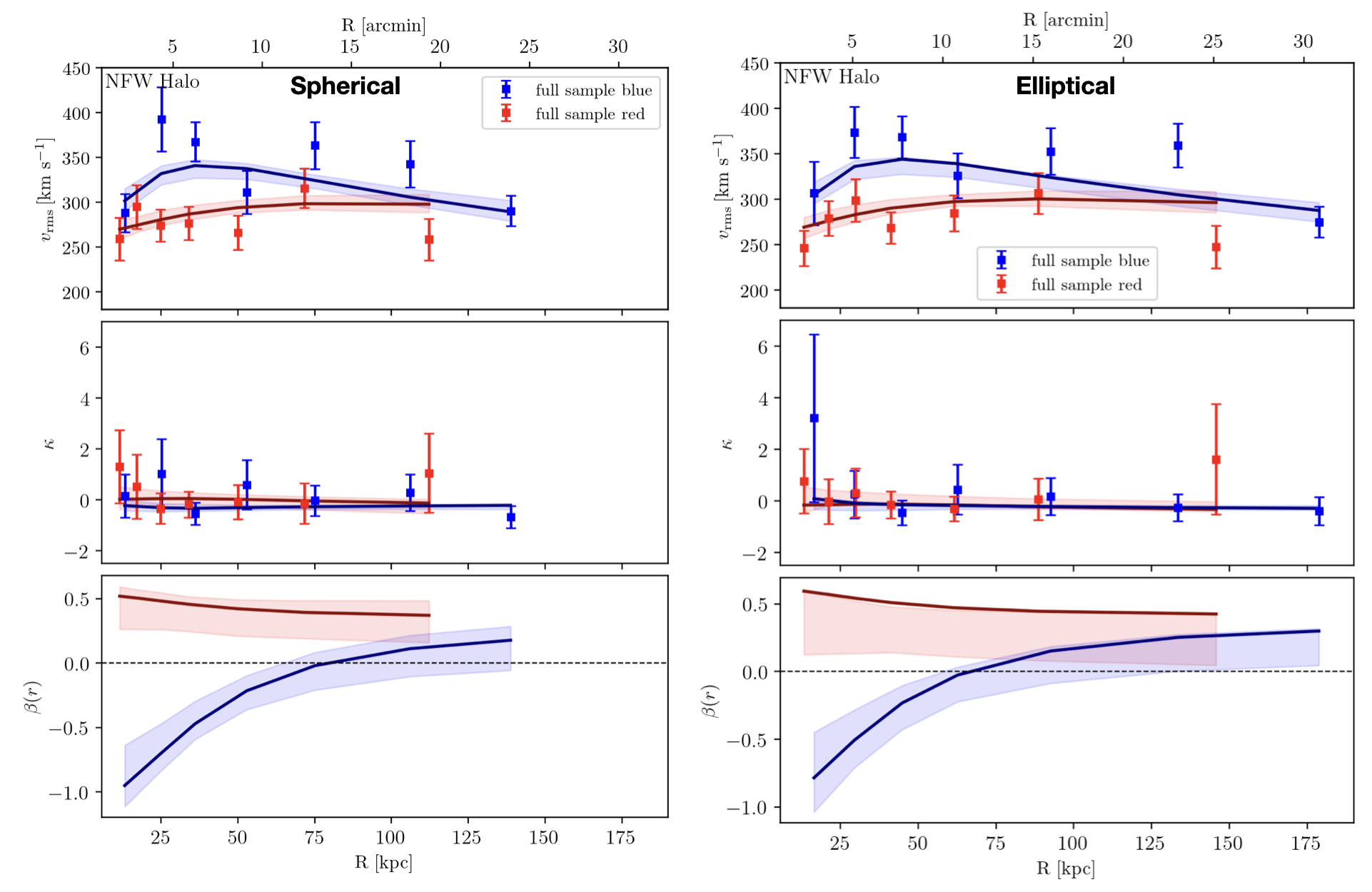}
    \caption{Observed and modelled velocity-dispersion and kurtosis profiles of the \textit{full} sample for the NFW halo. The left panels show the profiles extracted using circular annuli, while the right panels show the corresponding profiles extracted using elliptical annuli. The elliptical-annulus data points extend to larger radii because the radial coordinate corresponds to the semi-major axis of the ellipse. Although the binned kinematic profiles differ slightly between the two extractions, the spherical Jeans models provide comparably good fits in both cases.}
    \label{fig:c3}
\end{figure*}

\begin{figure*}
    \centering
    \includegraphics[width=0.45\linewidth]{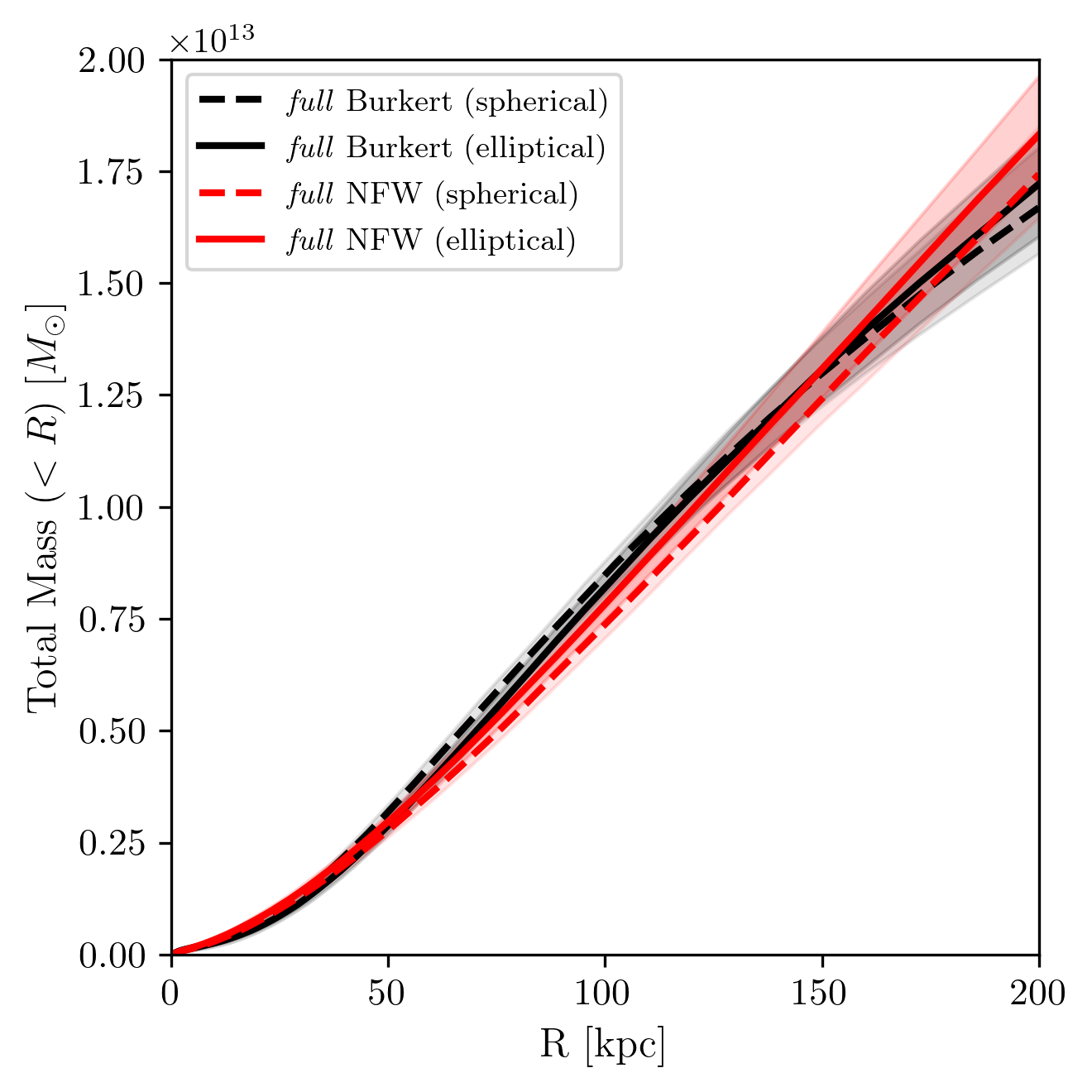}
    \caption{Enclosed mass profiles inferred from modelling the \textit{full} GC sample using tracer profiles extracted with circular and elliptical annuli. The dashed black and red curves show the results obtained from circular annuli for the Burkert and NFW halos, respectively, while the corresponding solid curves show the results obtained from elliptical annuli. The grey and red shaded regions indicate the 1$\sigma$ uncertainties for the Burkert and NFW mass profiles, respectively. The mass profiles derived from the two annulus choices agree within the 1$\sigma$ uncertainties, indicating that the assumed spherical geometry does not significantly affect the inferred enclosed mass profile.}
    \label{fig:c4}
\end{figure*}

\end{appendix}
\end{document}